\documentclass[aps,prd,showpacs,showkeys,floatfix,preprintnumbers,
footinbib]{revtex4}
\usepackage[utf8]{inputenc}
\usepackage{graphicx}
\usepackage{dcolumn}
\usepackage{bm}
\usepackage{amssymb}
\usepackage{amsmath}
\usepackage{subfigure}
\usepackage{tabularx}
\usepackage{hyperref}
\usepackage{color}
\usepackage{multirow,array}
\newcolumntype{P}[1]{>{\centering\arraybackslash}p{#1}}

\begin{document}
\title{Thermodynamics of strongly interacting matter in a hybrid model}
\author{Abhijit Bhattacharyya$^1$}
\email{abhattacharyyacu@gmail.com}
\author{Sanjay K. Ghosh$^2$}
\email{sanjay@jcbose.ac.in}
\author{Soumitra Maity$^2$}
\email{soumitra.maity1984@gmail.com}
\author{Sibaji Raha$^2$}
\email{sibaji@jcbose.ac.in}
\author{Rajarshi Ray$^2$}
\email{rajarshi@jcbose.ac.in}
\author{Kinkar Saha$^2$}
\email{saha.k.09@gmail.com}
\author{Subhasis Samanta$^3$}
\email{subhasis.samant@gmail.com}
\author{Sudipa Upadhaya$^2$}
\email{sudipa.09@gmail.com}
\affiliation{$^1$Department of Physics, University of Calcutta,\\
             92, A.P.C Road, Kolkata 700009, India}
\affiliation{$^2$Center for Astroparticle Physics \&
Space Science, \\ Block-EN, Sector-V, Salt Lake, Kolkata 700091, India 
 \\ \& \\ 
Department of Physics, Bose Institute,\\
93/1, A. P. C Road, Kolkata 700009, India}
\affiliation{$^3$School of Physical Sciences, National Institute of 
Science Education and Research, HBNI,\\ Jatni 752050,
India}
\begin{abstract}
The equation of state and fluctuations of conserved charges in a
strongly interacting medium under equilibrium conditions form the
baseline upon which various possible scenarios in relativistic heavy-ion
collision experiments are built. Many of these quantities have been
obtained in the lattice QCD framework with reliable continuum
extrapolations. Recently the Polyakov$-$Nambu$-$Jona-Lasinio model has
been reparametrized to some extent to reproduce quantitatively the
lattice QCD equation of state at vanishing chemical potentials. The
agreement was precise except at low temperatures, possibly due to
inadequate representation of the hadronic degrees of freedom in the
model. This disagreement was also observed for some of the fluctuation
and correlations considered. Here we address this issue by introducing
the effects of hadrons through the Hadron Resonance Gas model. The total
thermodynamic potential is now a weighted sum of the thermodynamic
potential of the Polyakov$-$Nambu$-$Jona-Lasinio model and that of the
Hadron Resonance Gas model. We find that the equation of state and the
fluctuations and correlations obtained in this hybrid model agrees
satisfactorily with the lattice QCD data in the low temperature regime.

\end{abstract}
\maketitle

\section{Introduction}
\label{sc.intro}

\par
The thermodynamic properties of strongly interacting matter under
extreme conditions are being actively investigated both theoretically as
well as experimentally. The lattice formulation of quantum
chromodynamics (QCD) on discretized space-time provides a first
principle approach in this direction~\cite{Karsch1}. In a system with
light quarks a rapid crossover from color confined and chiral symmetry
broken hadronic phase to a chirally restored and color deconfined
partonic phase has been predicted~\cite{Boyd, Engels, Katz, Hands,
Szabo, Laermann, Philipsen1, Aoki, Aoki1, yaoki, Katz1}. In the physical
case of two light and a heavier strange quark, lattice QCD simulations
at zero net conserved charges, this cross-over temperature is expected
to lie in the range $150~{\rm MeV} < T_c < 160~{\rm MeV}$ as reported by
the Hot-QCD~\cite{Bazavov12, Bazavov14} and
Wuppertal-Budapest~\cite{Borsanyi14} collaborations. Though at a
cross-over there is no singularity, it is observed that near $T_c$
various thermodynamic quantities exhibit a rapid
change~\cite{Hatta,Ejiri, Stephanov}. 

\par
At the same time, it is also important to properly explore the regions
of QCD phase diagram away from the temperature axis.  It is expected
that at some critical high baryon chemical potential and small
temperatures the system may undergo a first order phase transition from
hadronic to partonic phase~\cite{Asakawa:1989bq, Ejiri:2008xt,
Bowman:2008kc}. This first-order phase boundary would continue for some
lower chemical potentials and higher temperature, eventually terminating
at a critical end-point (CEP)~\cite{Halasz:1998qr, Fodor:2004nz,
Gavai:2004sd, Stephanov:2004wx}. One of the major goals in the
experiments with heavy ion collisions is to map this phase diagram of
QCD and locate the CEP, if any. 

\par
A reliable way to understand the phase transition dynamics is through
the study of correlations and fluctuations of conserved charges. At
finite temperatures and chemical potentials fluctuations of conserved
charges are sensitive indicators of the transition from hadronic matter
to partonic matter. Moreover, the existence of the CEP can be signalled
by the singularities in fluctuations. In lattice QCD framework many of
these susceptibilities have been obtained at vanishing chemical
potentials. Unfortunately for non-zero baryon chemical potentials the
lattice QCD framework faces certain technical difficulties. Recently
various intelligent techniques have been developed to circumvent these
difficulties to some extent~\cite{Katz, Finitemu2, Gavai:2001ie,
Laermann, Philipsen1, Finitemu1, Gavai1, Finitemu3, Allton1,
Bernard:2004je, Bernard:2007nm, Cheng08, Kaczmarek, Endrodi1, Endrodi2}.

\par
A parallel approach with QCD inspired models are being developed
alongside the lattice QCD approach to gain some insight into the various
aspects of strongly interacting matter. Here we shall discuss one such
model $-$ the Polyakov loop enhanced Nambu$-$Jona-Lasinio (PNJL) model.
Originally the PNJL model was introduced to enhance the
Nambu$-$Jona-Lasinio (NJL) model ~\cite{YNambu, Hatsuda2, Vogl,
Klevansky, Hatsuda1, Buballa1, Barducci} with the gluon thermodynamics
effectively through the Polyakov loop.  In the absence of the gluon
dynamics, the NJL model does not incorporate the mechanism of
confinement adequately.  Extending the NJL model to the PNJL model by
introducing a temporal background gluon-like field along with its self
interactions, restores some sense of confinement into the
model~\cite{Ogilvie, Fukushima, Ratti}.  The NJL as well as the PNJL
models conserve all the global charges like the chiral, baryon number,
electric charge, strangeness etc., as in QCD. The multi-quark
interactions in this model are responsible for the dynamical generation
of mass, leading to spontaneous breaking of chiral symmetry. 

\par
In the initial parametrizations of the PNJL model the NJL parameters
were set from the masses of observed hadron masses and decay
coefficients, while the Polyakov loop parameters were set from the pure
gauge dynamics on the lattice. This already led to qualitatively similar
results as in lattice QCD framework~\cite{Ratti, Ray, Mukherjee, Robner,
Ghosh, Datta, Fu, Wu}. Over the years several studies were done to
analyze the properties of this model as well as to improve the model
step by step. For example in order to stabilize the ground state of the
2+1 flavor system improvements were introduced in the PNJL
model~\cite{Bhattacharyya, Deb, Lahiri} following similar improvements
in the NJL model~\cite{Osipov, Hiller, Moreira, Blin}. A first case
study of the phase diagram in $\beta$-equilibrium using the PNJL model
was reported in~\cite{Majumder}. In a related work~\cite{Mustafa}, the
SU(3) color singlet ensemble of a quark-gluon gas has been shown to
exhibit a Z(3) symmetry and within stationary point approximation it
becomes equivalent to the Polyakov loop ensemble. Significant effects on
baryon-isospin correlations due to the mass difference between the two
light quarks were reported in~\cite{Raha}. Similarly the effects of
finite system sizes on various thermodynamic quantities including
fluctuations and correlations of conserved charges have been reported in
ref.~\cite{Sur, Bhatta}. Also the first model study of the net charge
fluctuations in terms of D-measure from the PNJL model~\cite{Kinkar} has
been reported. In fact the validity of the fluctuation-dissipation
theorem has also been discussed in the context of the PNJL
model~\cite{Anirban}. Viscous effects may play important role in the
evolution of the hot and dense system. Their effects in terms of
transport coefficients have been done in the NJL and PNJL
model~\cite{Redlich, Sabyasachi, Marty, Weise, Shi-Song, Sudipa, Saha,
Kaiser} and compared with hadron resonance gas studies~\cite{Das, Krein,
Kadam, Mohanty}. The behavioral pattern of various observables extracted
from the PNJL model may be found in~\cite{Claudia, Friman, Fuku, Kahara,
Zhang, Ruivo, Kashiwa, Buballa, Hansen, Lourenco, Majumder, Inagaki,
Friesen}, The QCD phase structure has also been investigated for
imaginary chemical potentials in the PNJL model framework~\cite{Sakai,
Yahiro, Morito}. Different interesting features of the Polyakov loop
have led to the development of different formalisms of the PNJL model
~\cite{Salcedo, Sal, Salc, Islam}. Effects of consideration of gluon
Polyakov loop have been discussed in ref.~\cite{Meg, Megi, Tsai}.
Recently improvements of the Polyakov loop potential have been carried
out by introducing the effects of back-reaction of the quarks
\cite{braun, rincon}.

\par
The qualitative agreement between the various observables computed in
the PNJL model and that in lattice QCD has been quite satisfactory. This
agreement seemed very convincing once the temperature dependent
observables were plotted against $T/T_c$, where $T_c$ in the model was
not equal to that obtained on the lattice. However lattice QCD data used
for these studies were only reported at finite lattice spacings.
Recently continuum extrapolations for a number of observables have been
reported from lattice simulations~\cite{Bazavov12, Bazavov14,
Bazavov:2012jq, Borsanyi:2011sw, Bazavov1213, Bazavov14a}.
Correspondingly the various thermodynamic properties of strongly
interacting matter were investigated within the framework of the PNJL
model by reparametrizing the Polyakov loop potential~\cite{repara} .
The focus was to ascertain a quantitative agreement of a variety of
observables with the lattice data. The overall correspondence was
satisfactory but not perfect. One of the regions of disagreement was in
the low temperature region where hadronic degrees of freedom are
dominant. This may be expected as these degrees of freedom are not
adequately addressed in the PNJL model. 

\par
On the other hand, the hadron resonance gas (HRG)
model~\cite{HRG_Braun-Munzinger} has been very successful in describing
the hadron yields in central heavy ion collisions from AGS up to RHIC
energies~\cite{BraunMunzinger:1994xr, 9603004_Cleymans,
PLB465_Braun-Munzinge, PRC60_054908_Cleymans, PLB518_Braun-Munzinger,
Xu:2001zj, PRC73_Becattini, NPA772_Andronic, Cleymans:2005xv,
Andronic:2008gu, Andronic:2009jd, Karsch:2010ck, Chatterjee:2015fua}.
Along with that the susceptibilities of conserved charges calculated in
lattice QCD have been well reproduced by HRG model \cite{Karsch:2003zq,
Tawfik:2004sw, Andronic:2012ut, Bhattacharyya:2013oya} for temperatures
up to 150 MeV. Also the region of large chemical potentials below the
critical region can be studied using this model. Thus this model is
quite suitable for describing the hadronic phase of strong interactions.
The HRG model is based on the Dashen, Ma and Bernstein
theorem~\cite{dashen} which shows that a dilute system of strongly
interacting matter can be described by a gas of free resonances. Though
the long range attractive part of the hadron interactions is taken care
of by these resonances the short range repulsion are also important for
the description of strongly interacting matter. Near the
critical/cross-over region HRG calculations tend towards Hagedorn
divergence which may be due to the absence of repulsive interaction.
This repulsive part is incorporated through the excluded volume effects
in the HRG and is commonly known as EVHRG model~\cite{hagedron:1980,
ZPC51_Rischke, Cleymans:1992jz, Singh:1991np, Yen:1997rv,
PRC77_Gorenstein, Andronic:2012ut, PRC85_Fu, begunprc88, PLB722_Fu,
PRC88_Tawfik}. EVHRG equation of states have also been used for the
hydrodynamic models of nucleus-nucleus collision~\cite{hama, werner,
satarov}. Recently fluctuations of conserved charges, using HRG model,
have been studied in Ref.~\cite{Garg:2013ata} including the experimental
acceptances of pseudo-rapidity and transverse momentum. In
Ref.~\cite{PLB722_Fu}, higher moments of net-proton multiplicity have
been studied using EVHRG model and compared with STAR data. A similar
study was done by some of us for net-proton, net-charge and net-kaon
using the HRG and EVHRG models~\cite{Bhattacharyya:2013oya}. The effects
of finite volumes on the hadron gas were discussed in
\cite{Bhattacharyya:2015zka} and effects of magnetic fields were
presented in \cite{hrgmag}. In fact as the higher moments are expected
to be more sensitive to the phase transition, any deviation of
experimental observation from the model results may be taken as an
indication of new phenomena. Investigation in this direction was done
with variation of collision centrality as well as collision energy in
\cite{hrgcentral}.

\par
Unfortunately, there is no single model which describes both the low
temperature and high temperature domain accurately. However, a possible
mechanism to address this issue has been suggested
in~\cite{Albright:2014gva}. The idea is to make an interface between one
model appropriate for the hadronic phase and another model appropriate
for the partonic phase.  For the hadronic sector the authors have used
the HRG model, and for the partonic sector they have used the PQCD
results, and have used a switching function at the interface.  In a
follow up work~\cite{Albright15} they have discussed the various
versions of HRG models along with the PQCD data and obtained quite a
number of thermodynamic variables including some susceptibilities in the
baryon sector. They found the equation of state to be reproduced very
well vis-a-vis the lattice QCD data, but the values of some of the
susceptibilities deviate slightly from the lattice QCD data.  However
the baryonic susceptibilities obtained in STAR data are well reproduced
with the EVHRG model for the hadronic sector~\cite{Albright16}. 

\par
In a similar study an interacting quark (IQ) model, which is a modified
PNJL model was used for the partonic phase in~\cite{iq-hrg1, iq-hrg2},
and HRG model for the hadronic phase. Here the multiquark interactions
in the NJL sector have been neglected and the quark masses are taken to
be the nominal Lagrangian masses. In these studies the equation of state
and susceptibilities of various conserved charges have been computed and
reproduces the lattice QCD data quite well.

\par
In the present study we shall explore this approach utilizing the PNJL
model and the HRG model. Before going on to the details let us clarify
some of our assumptions and some of the differences with the earlier
studies. Here we shall consider the simplest non-interacting form of the
HRG model to describe the low temperature phase. We expect the effects
of the short range repulsive forces incorporated in the excluded volume
HRG model to be small for $T<150$ Mev~\cite{Bhattacharyya:2013oya}.  As
we shall see, the non-interacting HRG model easily describes most of the
physics in the low temperature phase.

\par
Secondly, in both sets of the earlier studies the switching function is
considered to have both temperature and chemical potential dependence.
This should be true in general. However for simplicity we have chosen it
to be temperature dependent as in the present work we shall only
concentrate on observables at zero chemical potential.

\par
Lastly, all the earlier studies have considered derivatives of the
switching function for computing further observables using the
thermodynamic relations. For example if switching function is used as an
interpolating function of hadronic and partonic pressure, the entropy
will be obtained from the temperature derivative, which also acts on the
switching function. We however assume the switching function has no
thermodynamic origin. It is used as a global interpolator between the
hadronic and partonic phases irrespective of the observables. So
thermodynamic operators should not act on the switching function.
Nonetheless, we have discussed what happens if the temperature
derivatives of switching function are included for some of the
thermodynamic observables. As we discuss below that with proper
parametrizations the numerical results do not depend much on whether the
derivatives are included or not. Similar derivatives of the switching
functions were considered in the IQ-HRG approach~\cite{iq-hrg2}, where
the derivative terms were used as parameters to fit the lattice QCD data
for entropy and susceptibilities. Since we shall not consider the
chemical potential dependence of the switching function, these
derivative effects in the susceptibilities in our study are zero. Yet we
shall see that the results seem to satisfactorily reproduce lattice QCD
data in the hadronic phase.

\par
We have organized the manuscript as follows. In the next two
sections~\ref{sc.pnjl} and~\ref{sc.hrg}, we shall briefly discuss the
PNJL model and the HRG model respectively. In section~\ref{sc.switch} we
discuss how to couple the PNJL and HRG models, utilizing some of the
thermodynamic quantities relating to the equation of state. Thereafter
in section \ref{sc.fluccor} we discuss the predicted behavior of the
fluctuations and correlations of conserved charges from the PNJL-HRG
model. And finally in section~\ref{sc.conclusion} we summarize and
conclude.

\section{PNJL Model}
\label{sc.pnjl}

In this section we briefly discuss the PNJL model used in this study. A
detailed discussion may be found in Ref.~\cite{repara}.
The PNJL model was initialized  with a Polyakov loop effective potential
being added to the NJL model~\cite{Ogilvie, Fukushima, Ratti}. While the
chiral properties are taken care of by the NJL part, the Polyakov loop
explains the deconfinement physics. Extensive studies have been carried
out using PNJL model with 2 and 2+1 flavors~\cite{Ratti, Ray, Mukherjee,
Ghosh, ciminale, Bhattacharyya, Shao, Tang, Haque, Peixoto}. Here we
consider 2+1 flavor PNJL model taking up to six and eight quark
interaction terms as in ~\cite{ciminale, Bhattacharyya}. The
thermodynamic potential in this case reads as,

\small
\begin{eqnarray}
\Omega (\Phi,\bar{\Phi},\sigma_f,T,\mu ) &=&
2g_S\sum_{f=u,d,s}\sigma_f^2 -
\frac{g_D}{2}\sigma_u\sigma_d\sigma_s + 
3\frac{g_1}{2}(\sum_f\sigma_f^2)^2 + 3g_2\sum_f\sigma_f^4
- 6\sum_f\int_0^\infty\frac{d^3p}{(2\pi)^3}E_f
\Theta(\Lambda-|\vec{p}|)
\nonumber \\
&& - ~2T\sum_f\int_0^\infty\frac{d^3p}{(2\pi)^3} \ln{\left[ 
1+3\left( \Phi+\bar{\Phi} e^{-\left( E_f-\mu_f \right) /T} \right)
e^{-\left( E_f-\mu_f \right) /T } + 
e^{-3\left( E_f-\mu_f \right) /T } \right]  }
\nonumber \\
&& - ~2T\sum_f\int_0^\infty\frac{d^3p}{(2\pi)^3} \ln{\left[ 
1+3\left( \bar{\Phi}+\Phi e^{-\left( E_f+\mu_f \right) /T} \right)
e^{-\left( E_f+\mu_f \right) /T } + 
e^{-3\left( E_f+\mu_f \right) /T } \right]  }
\nonumber \\
&&
+ ~\mathcal{U'}(\Phi,\bar{\Phi},T)
\label{eq.Potential}
\end{eqnarray}
\normalsize

\noindent
The fields $\sigma_f=\langle\bar\psi_f\psi_f\rangle$ correspond to the
two light flavor ($f=u,d$) condensates and the strange ($f=s$) quark
condensate respectively. The model has a four quark coupling term with
coefficient $g_S$, a six quark coupling term breaking the axial U(1)
symmetry explicitly with a coefficient $g_D$, and eight quark coupling
terms with coefficients $g_1$ and $g_2$ necessary to sustain a stable
minima in the NJL Lagrangian.  The corresponding quasiparticle energy
for a given flavor $f$ is $E_f=\sqrt{p^2+M_f^2}$, with the dynamically
generated constituent quark masses given by,

\begin{equation}
M_f=m_f-2g_S\sigma_f+\frac{g_D}{2}\sigma_{f+1}\sigma_{f+2}-2g_1\sigma_f
(\sigma_u^2+\sigma_d^2+\sigma_s^2)-4g_2\sigma_f^3
\end{equation} 

\noindent
In the above, if $\sigma_f=\sigma_u$, then $\sigma_{f+1}=\sigma_d$ and
$\sigma_{f+2}=\sigma_s$, and so on in a clockwise manner.  The finite
range integral gives the zero point energy. The different parameters as
obtained from ~\cite{Bhattacharyya} are given in Table~\ref{tb.njlpara}.

\begin{table}[!htb]
\begin{tabular}{|c|c|c|c|c|c|c|c|}
\hline
\hline
Interaction & $m_u$ (MeV) & $m_s$ (MeV) & $\Lambda$ (MeV) &
$g_s\Lambda^2$ & $g_D\Lambda^5$ & $g_1\times 10^{-21}$ (MeV$^{-8}$) &
$g_2\times 10^{-22}$ (MeV$^{-8}$) \\
\hline
\hline
6-quark &  5.5 & 134.758 & 631.357 & 3.664 & 74.636 & 0.0  & 0.0 \\
\hline
8-quark & 5.5  & 183.468 & 637.720 & 2.914 & 75.968 & 2.193 & -5.890 \\
\hline
\hline
\end{tabular}
\caption{Parameters in the NJL model}
\label{tb.njlpara}
\end{table}

\noindent
The finite temperature and chemical potential contributions of the
constituent quarks are given by the next two terms. Note that these are
basically coming from the fermion determinant in the NJL model modified
due to the presence of the fields corresponding to the traces of
Polyakov loop and its conjugate given by $\Phi=\frac{Tr_cL}{N_c}$ and
$\bar{\Phi}=\frac{Tr_cL^{\dagger}}{N_c}$ respectively. Here
$L(\vec{x})=\mathcal{P} exp\left[ i\int_0^{1/T}d\tau A_4(\vec{x},\tau)
\right]$ is the Polyakov loop, and $A_4$ is the temporal component of
background gluon field.

\par
The effective potential that describes the self interaction of the
$\Phi$ and $\bar{\Phi}$ fields are given by $\mathcal{U'}$. Various
forms of the potential exist in the literature (see e.g.
~\cite{Robner, Fuku, Ghosh, Contrera, Qin}). We shall use the form
prescribed in~\cite{Ghosh} which reads as,

\begin{equation}
\frac{\mathcal{U'}(\Phi,\bar{\Phi},T)}{T^4}=
\frac{\mathcal{U}(\Phi,\bar{\Phi},T)}{T^4}-\kappa
ln[J(\Phi,\bar{\Phi})].
\label{eq.Ppotential}
\end{equation}

\noindent
Here ${\mathcal U}(\Phi,\bar{\Phi},T)$ is a Landau-Ginzburg type
potential commensurate with the global Z(3) symmetry of the Polyakov
loop~\cite{Ratti}. $J(\Phi,\bar{\Phi})$ is the Jacobian of
transformation from the Polyakov loop to its traces, and $\kappa$ is a
dimensionless parameter which is determined phenomenologically.  The
effective potential is chosen to be of the form,

\begin{equation}
\frac{\mathcal{U}(\Phi,\bar{\Phi},T)}{T^4}=
-\frac{b_2(T)}{2}\bar{\Phi}\Phi-\frac{b_3}{6}
(\Phi^3+\bar{\Phi}^3)+\frac{b_4}{4}(\bar{\Phi}\Phi)^2
\end{equation}

\noindent
The coefficient $b_2(T)$ is chosen to have a temperature dependence of
the form,

\begin{equation}
b_2(T)=a_0+a_1exp(-a_2\frac{T}{T_0})\frac{T_0}{T},
\end{equation}

\noindent
and $b_3$ and $b_4$ are chosen to be constants.
The set of parameters is given in Table~\ref{tb.polpara}.

\begin{table}[!htb]
\begin{tabular}{|c|c|c|c|c|c|c|c|}
\hline
\hline
Interaction & $T_0$ (MeV) & $a_0$ & $a_1$ & $a_2$ & $b_3$ & $b_4$ &
$\kappa$\\
\hline
\hline
6-quark & 175 & 6.75 & -9.0 & 0.25 & 0.805 & 7.555 & 0.1 \\
\hline
8-quark & 175 & 6.75 & -9.8 & 0.26 & 0.805 & 7.555 & 0.1    \\
\hline \hline
\end{tabular}
\caption{Parameters for the Polyakov loop potential.}
\label{tb.polpara}
\end{table}

\begin{table}[!htb]
\begin{tabular}{|c|c|c|c|}
\hline
\hline
Interaction & Peak position of $d\Phi/dT$ (MeV) &
Peak position of $d\sigma/dT$ (MeV) & $T_c$ (MeV) \\
\hline
\hline
6-quark & 142 & 191 & 166.5 \\
\hline
8-quark & 158 & 167 & 162.5 \\
\hline \hline
\end{tabular}
\caption{Location of crossover temperature}
\label{tb.tc}
\end{table}

\par
Using realistic quark masses the deconfinement temperature obtained in
lattice QCD is found to be much higher than the chiral transition
temperature \cite{Aoki, yaoki}. However the deconfinement temperature as
measured from the peak of the entropy of a static quark is found to be
consistent with the chiral transition temperature \cite{jweber}. In our
model framework we consider the temperature derivatives of the mean
fields and locate their peaks to obtain the transition temperature. The
temperature derivative of the Polyakov loop is closely related to the
definition of temperature derivative of the static quark free energy
that gives its entropy as defined in \cite{jweber}. The corresponding
$T_c$ was obtained from the average of the two peak positions. The
resulting values of $T_c$ are listed in Table \ref{tb.tc}.

\par
With this parametrization we have been able to achieve both a crossover
temperature of $T_c\sim160~{\rm MeV}$ as well as quantitative agreement
of temperature variation of pressure and various other observables with
the lattice QCD continuum estimation~\cite{repara}. However the
quantitative agreement though close, was not precise enough. Various
observables showed disagreement in different ranges of temperatures. The
most common discrepancies were found in the low temperature region where
the hadronic degrees of freedom dominate. We shall thus make an attempt
to remove this lacunae by coupling the PNJL model with the HRG model.

\section{Hadron resonance gas model}
\label{sc.hrg}

We now discuss the HRG model briefly. The detailed discussions may be
found in Refs.~\cite{HRG_Braun-Munzinger, BraunMunzinger:1994xr,
9603004_Cleymans, PLB465_Braun-Munzinge, PRC60_054908_Cleymans,
PRC73_Becattini, PLB518_Braun-Munzinger, NPA772_Andronic,
Andronic:2008gu, hagedron:1980, ZPC51_Rischke, Cleymans:1992jz,
Singh:1991np, Yen:1997rv, Andronic:2012ut}. The grand canonical
partition function of a hadron resonance gas~\cite{HRG_Braun-Munzinger,
Andronic:2012ut} can be written as, 

\begin {equation}
 \ln Z^{id}=\sum_i \ln Z_i^{id},
\end{equation}

\noindent
where the sum is over all the hadrons. $id$ refers to an ideal gas of
hadronic resonances. The partition function for the $i^{th}$ resonance
is,

\begin{equation}
 \ln Z_i^{id}=\pm \frac{Vg_i}{2\pi^2}\int_0^\infty p^2\,dp
\ln[1\pm\exp(-(E_i-\mu_i)/T)],
\end{equation}

\noindent
where $V$ is the volume of the system, $g_i$ is the degeneracy factor,
$T$ is the temperature, $E_i=\sqrt{{p}^2+m^2_i}$ is the single particle
energy, $m_i$ is the mass and $\mu_i=B_i\mu_B+S_i\mu_S+Q_i\mu_Q$ is the
chemical potential. $B_i,S_i,Q_i$ are respectively the baryon number,
strangeness and charge of the particle, $\mu^,s$ being corresponding
chemical potentials. The $(+)$ and $(-)$ sign corresponds to fermions
and bosons respectively.  From the partition function one can calculate
various thermodynamic observables of the system created in heavy ion
collisions. In this work we shall incorporate all the hadrons listed by
the Particle Data Group \cite{pdg} up to mass of 3 GeV.

\section{Coupling the hadronic and partonic sectors}
\label{sc.switch}

\par
Let us now discuss the framework in which the hadronic matter described
using the HRG model can be smoothly {\it switched} to partonic matter
modelled through the PNJL one. The basic procedure is similar to the
ones reported in~\cite{Albright:2014gva, Albright15, Albright16,
iq-hrg1, iq-hrg2}. The pressure of the system as taken to be a sum of
partial pressures of the hadronic and partonic matter, weighted with a
switching function as,

\begin{equation}
P(T)=S(T)P_P(T) + (1-S(T))P_H(T),
\label{pressureeq}
\end{equation}

\noindent
where $P_P(T)$ and $P_H(T)$ are the pressures of partonic and hadronic
sectors respectively and $S(T)$ is the switching function. For the
partonic pressure we shall make a comparative study between PNJL models
with six and eight quark type interactions respectively. 

\begin{figure}[!htb]
\includegraphics[scale=0.28,angle=270]{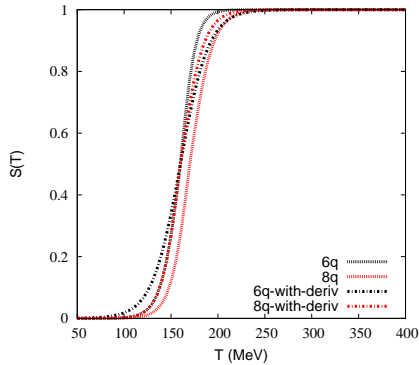}
\caption{(Color online) Switching function as a function of
temperature.}
\label{fg.switch}
\end{figure}

\par
Ideally $S(T)$ should be zero in the hadronic phase and $1$ in the
partonic phase. However since there is no singular phase boundary
separating the two phases, we consider the switching function to
interpolate smoothly from $0$ at low temperatures to $1$ at high
temperatures. We have assumed the switching function to be independent
of chemical potential at zero chemical potentials. The functional form
of the switching function is given as,

\begin{equation}
S(T)=\frac { 1 }{ 1+exp\left[ -\cfrac{T-T_S}{\Delta T_S(T)} \right]  }.
\end{equation}

\noindent
Here $T_S$ and $\Delta T_S(T)$ are parameters whose values should be
closely related to the crossover temperature and width of the crossover
region respectively. Note that the form of the switching function is
somewhat different from the earlier studies. The essential difference is
the introduction of two temperature scales $-$ one being the central
value $T_S$ of the switching and the other being the spread $\Delta
T_S$. $T_S$ is related to the cross-over temperature and $\Delta T_S(T)$
to the spread of the switching function on the temperature axis from its
central value. Given that the cross-over temperature is not sharply
defined, we have let $T_S$ to be fixed at $160$ MeV.  Thereafter we try
to find the best possible value of $\Delta T_S(T)$ so as to obtain the
closest quantitative agreement of pressure as obtained in lattice QCD.

\par
The procedure for obtaining the various thermodynamic observables are as
follows. We first obtain the pressure in the HRG model as well as the
PNJL model from the respective thermodynamic potentials. Thereafter we
obtain the hybrid pressure according to Eq.~(\ref{pressureeq}).  Once we
have the pressure the other thermodynamic quantities like entropy,
specific heat etc., are obtained from the temperature derivatives of the
hybrid pressure.  In all the previous studies the temperature and
chemical potential derivatives of the switching function in the hybrid
pressure were included as well. We however consider the switching
function to be defined globally for all the observables considered, so
that the effective function should be independent of the observable
being measured. But we shall still make a comparative study by both
including and excluding the temperature derivative terms for some of the
thermodynamic variables.

\par
The entropy density is obtained from the first derivative of pressure
with respect to temperature and is given by,

\begin{equation}
s(T) = \cfrac{\partial P}{\partial T} =
[S(T)s_P(T)+(1-S(T))s_H(T)]+(P_P(T)-P_H(T))
\cfrac{\partial S(T)}{\partial T} 
\end{equation}

\noindent
Here $s_P(T)$ and $s_H(T)$ are the entropy density for the PNJL model
and HRG model respectively. Note that the term in the square brackets
denote the hybrid entropy due to the switching function, while the
other term denotes the effects of the derivatives of the switching
function itself. Similar would be the case for any other thermodynamic
quantity where temperature derivative is involved. For example the
energy density is given by,

\begin{equation}
\epsilon(T) = Ts(T)-P(T) =
[S(T)\epsilon_P(T)+(1-S(T))\epsilon_H(T)]+T(P_P(T)-P_H(T))
\cfrac{\partial S(T)}{\partial T}. 
\end{equation}

\noindent
Here again the term within the square brackets would give the hybrid
energy density and the other term comes from the derivative of $S(T)$.
The derivative of $S(T)$ is given as,

\begin{equation}
\cfrac{\partial S(T)}{\partial T} =
\frac{1}{\Delta T_S} \left[ S(T)-S(T)^2 \right]  
\end{equation}

\par
The derivatives of $S(T)$ contribute more terms for higher order
temperature derivatives. For the specific heat at constant volume we get
the expression as,

\begin{eqnarray}\nonumber
C_V(T)&=& \left(\cfrac{\partial \epsilon}{\partial T}\right)_V= 
[S(T){C_V}_P(T)+(1-S(T)){C_V}_H(T)]\nonumber \\
&& +2T(s_P(T)-s_H(T))\cfrac{\partial S(T)}{\partial T}
+T(P_P(T)-P_H(T))\cfrac{\partial^2 S(T)}{\partial T^2}.
\end{eqnarray}

\noindent
Here the second order derivative term of $S(T)$ is given as,
\begin{equation*}
\frac{{\partial}^2 S(T)}{\partial T^2} =
\frac{1}{{\left( \Delta T_S \right)}^2}
\left[S(T)-3S(T)^2+2S(T)^3 \right].
\end{equation*}

\begin{figure}[!htb]
\centering
\includegraphics[scale=0.28,angle=270]{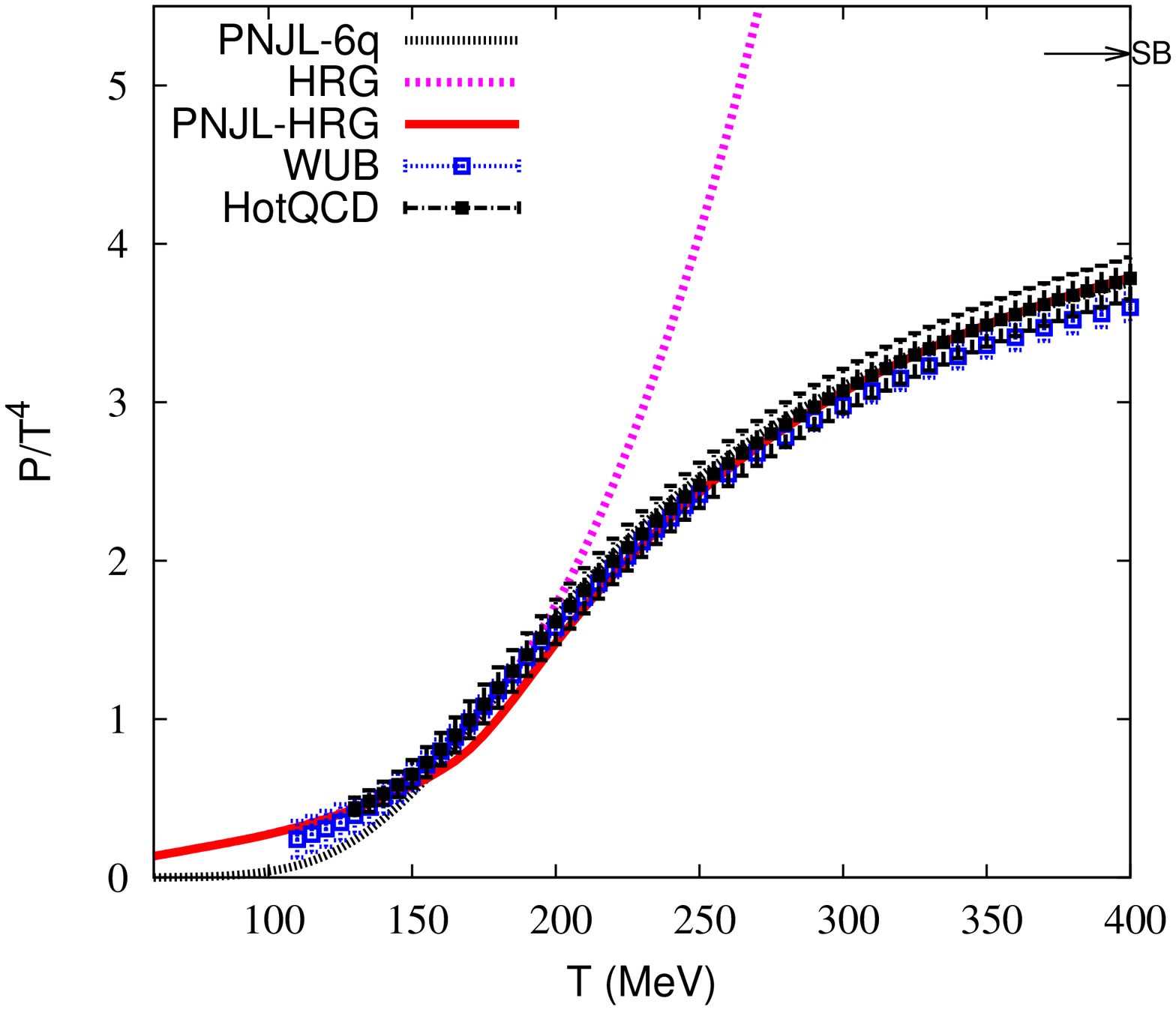}
\includegraphics[scale=0.28,angle=270]{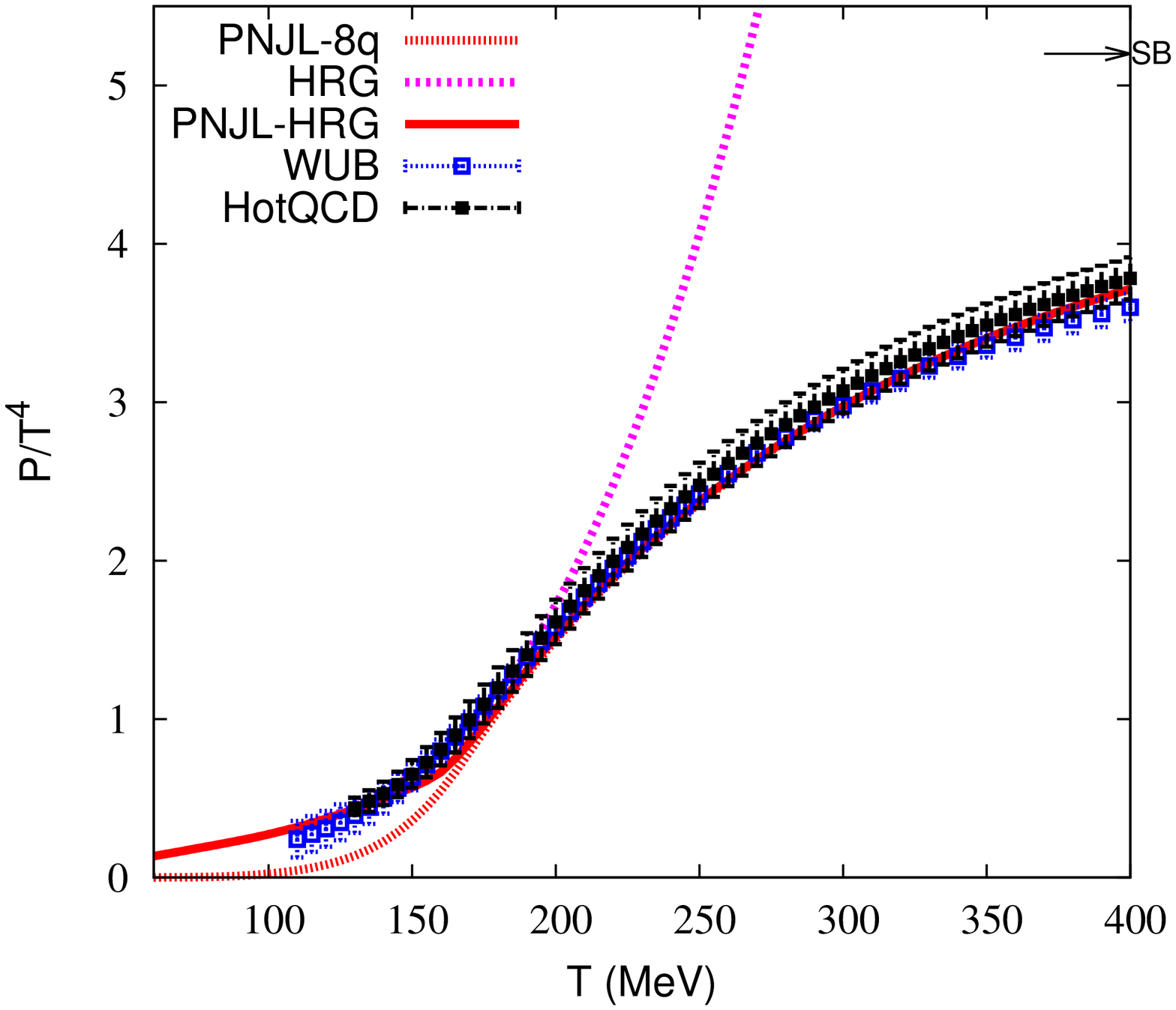}\\
\includegraphics[scale=0.28,angle=270]{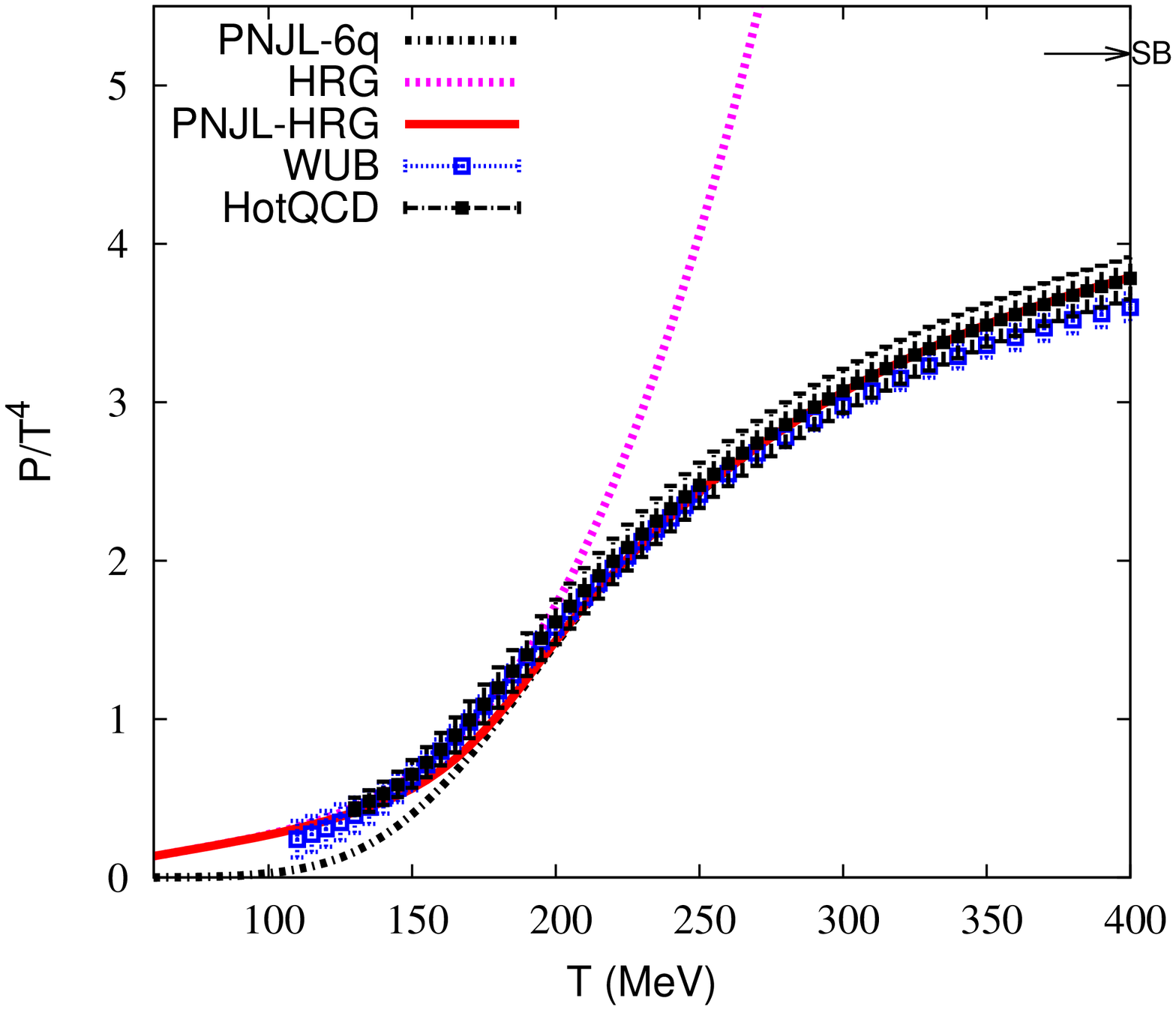}
\includegraphics[scale=0.28,angle=270]{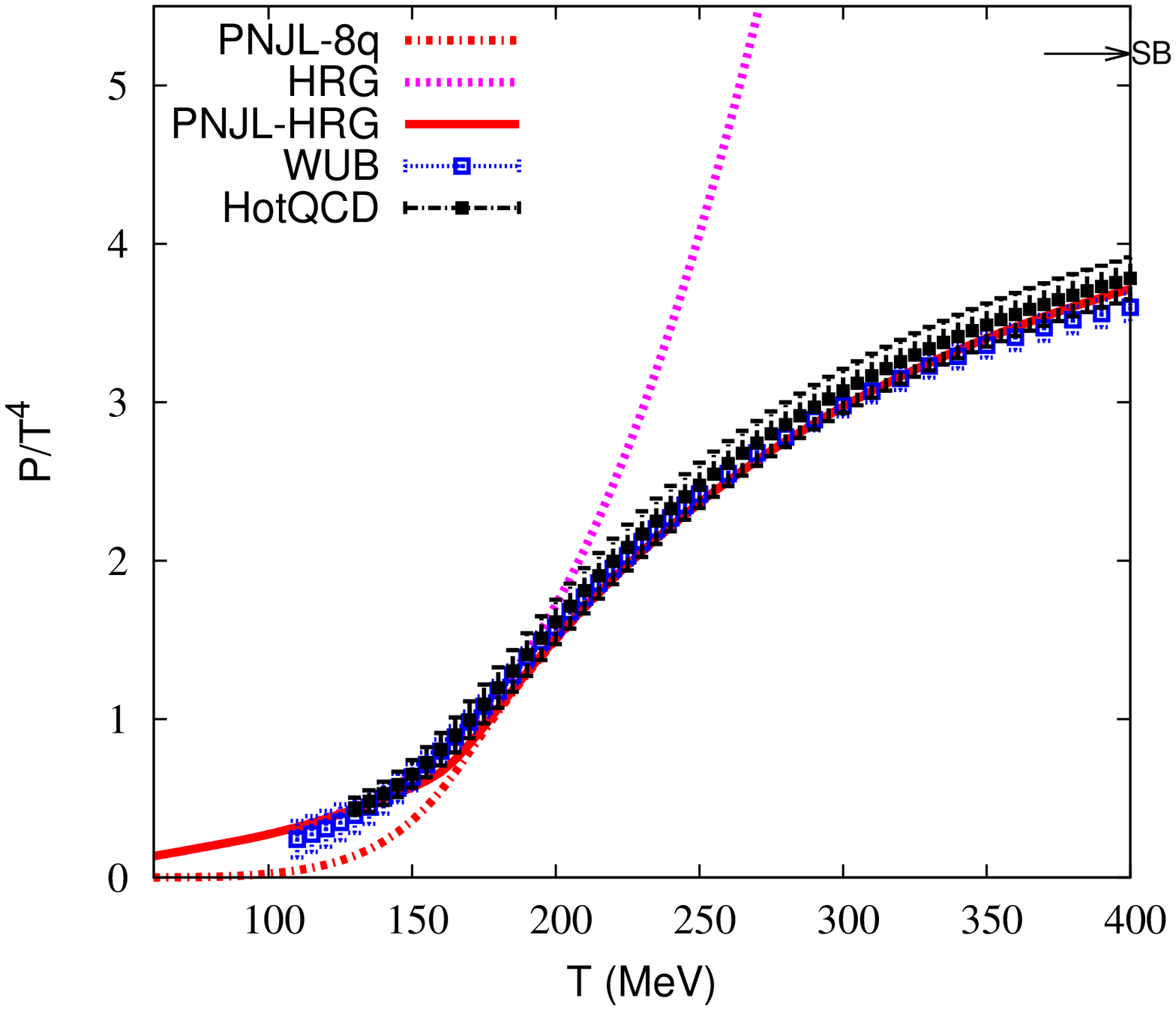}
\caption{(Color online) Scaled pressure as a function of temperature.
The upper panels are drawn without derivatives of switching function and
lower panels include those derivatives.  The left panel is for PNJL
model with six quark interactions and the right panel for PNJL model
with eight quark interactions. The continuum extrapolated lattice QCD
data are taken from Ref.~\cite{Bazavov14} (HotQCD) and
Ref.~\cite{Borsanyi14} (WUB).} \label{fig.pressure} \end{figure}

\par
We now show the temperature variations of some of the thermodynamic
quantities. We shall compare the results in PNJL model with six and
eight quark interactions respectively. We shall also compare results
including and excluding the temperature derivatives of the switching
function. As mentioned earlier the only parameter adjusted in the
switching function is $\Delta T_S(T)$. It is allowed to take two
different values on the two sides of $T_S$ when the derivatives of
$S(T)$ are excluded. Here the values of $\Delta T_S(T)$ are not
uniquely determined by any standard fitting procedure.  However for the
purpose of this paper it suffices to choose the values that best
describe most of the thermodynamic quantities in the hybrid systems by
trial and error. These values of $\Delta T_S(T)$ are given in Table
(\ref{tb.deltat0}). Results are not much affected by shifting these
values by $10\%$. The form of the switching function is shown in
Fig.~(\ref{fg.switch}). 

\begin{table}[!htb]
\begin{tabular}{|c|P{4.2 cm}|P{2.1 cm}|P{2.1 cm}|}
\hline
\hline
\multirow{2}{*}{PNJL Interactions} & 
\multicolumn{1}{P{4.2 cm}|}{$\Delta T_S$ (MeV) \hskip 1 in 
Including derivatives of S(T)} &
\multicolumn{2}{P{4.2 cm}|}{$\Delta T_S$ (MeV) \hskip 1 in
Excluding derivatives of S(T)} \\
\cline{2-4}
 & For all values of T & $T<T_S$ & $T>T_S$\\
\hline
\hline
6q & 15 & 10 & 8\\
\hline
8q & 15 & 10 & 18\\
\hline
\hline
\end{tabular}
\caption{Values of $\Delta T_S$ used.}
\label{tb.deltat0}
\end{table}

\par
In Fig. (\ref{fig.pressure}) we show scaled pressure as a function of
temperature for various models and compare them to the lattice QCD data
~\cite{Bazavov14, Borsanyi14}. While the HRG model agrees with lattice
QCD data below $T=200$ MeV, the PNJL model agrees with lattice QCD data
above $T=150$ MeV. On the other hand the combined PNJL-HRG model now
agrees with the lattice QCD data for the full temperature range. The
differences in the results due to the type of PNJL model chosen are
quite insignificant. However some of those effects are absorbed in the
values of $\Delta T_S$ as given in table~(\ref{tb.deltat0}). We note
here that the pressure itself only depends on $S(T)$ and not on any of
its derivatives. However the other thermodynamic variables are obtained
from the derivatives of the hybrid pressure. As mentioned earlier, we
are choosing $\Delta T_S$ such that the bulk thermodynamic quantities
obtained in lattice QCD framework are well reproduced in the PNJL-HRG
model.  Thus the values of pressure would depend on which form of $S(T)$
we are choosing. This is why we also show the plot for pressure for
those values of $\Delta T_S$ that are chosen when derivatives of $S(T)$
are included in computing rest of the thermodynamic variables.

\begin{figure}[!htb]
\centering
\includegraphics[scale=0.28,angle=270]{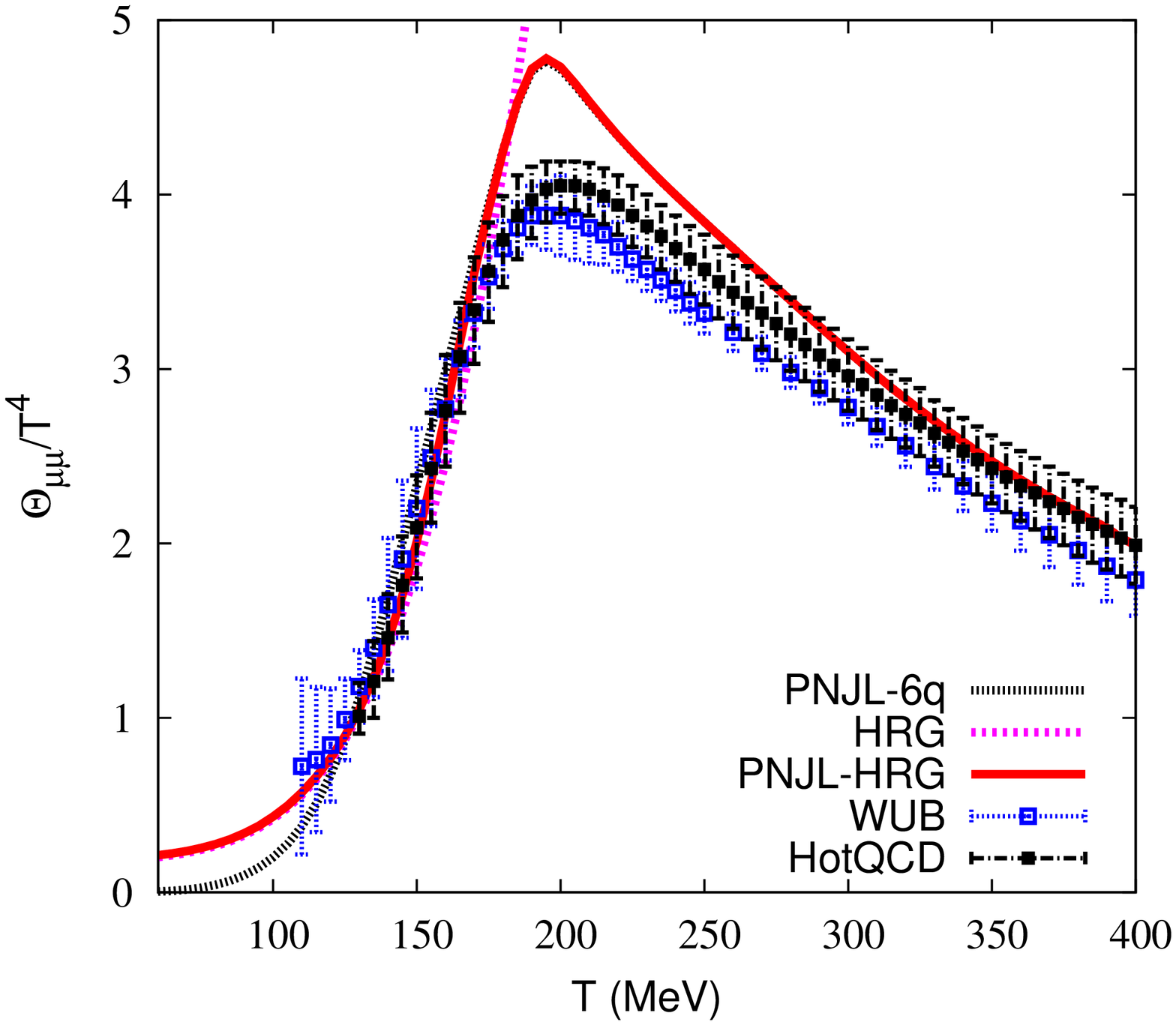}
\includegraphics[scale=0.28,angle=270]{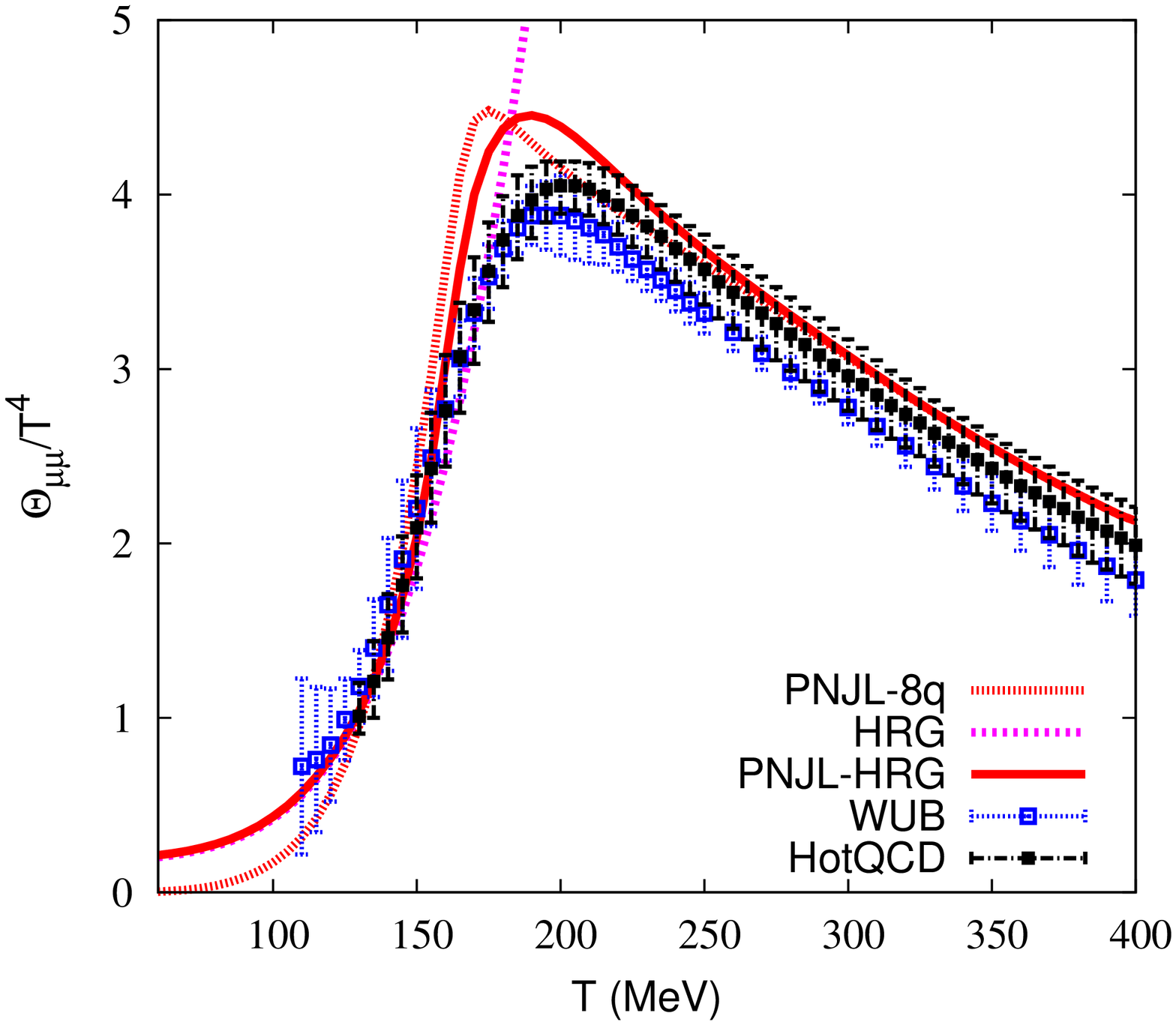}\\
\includegraphics[scale=0.28,angle=270]{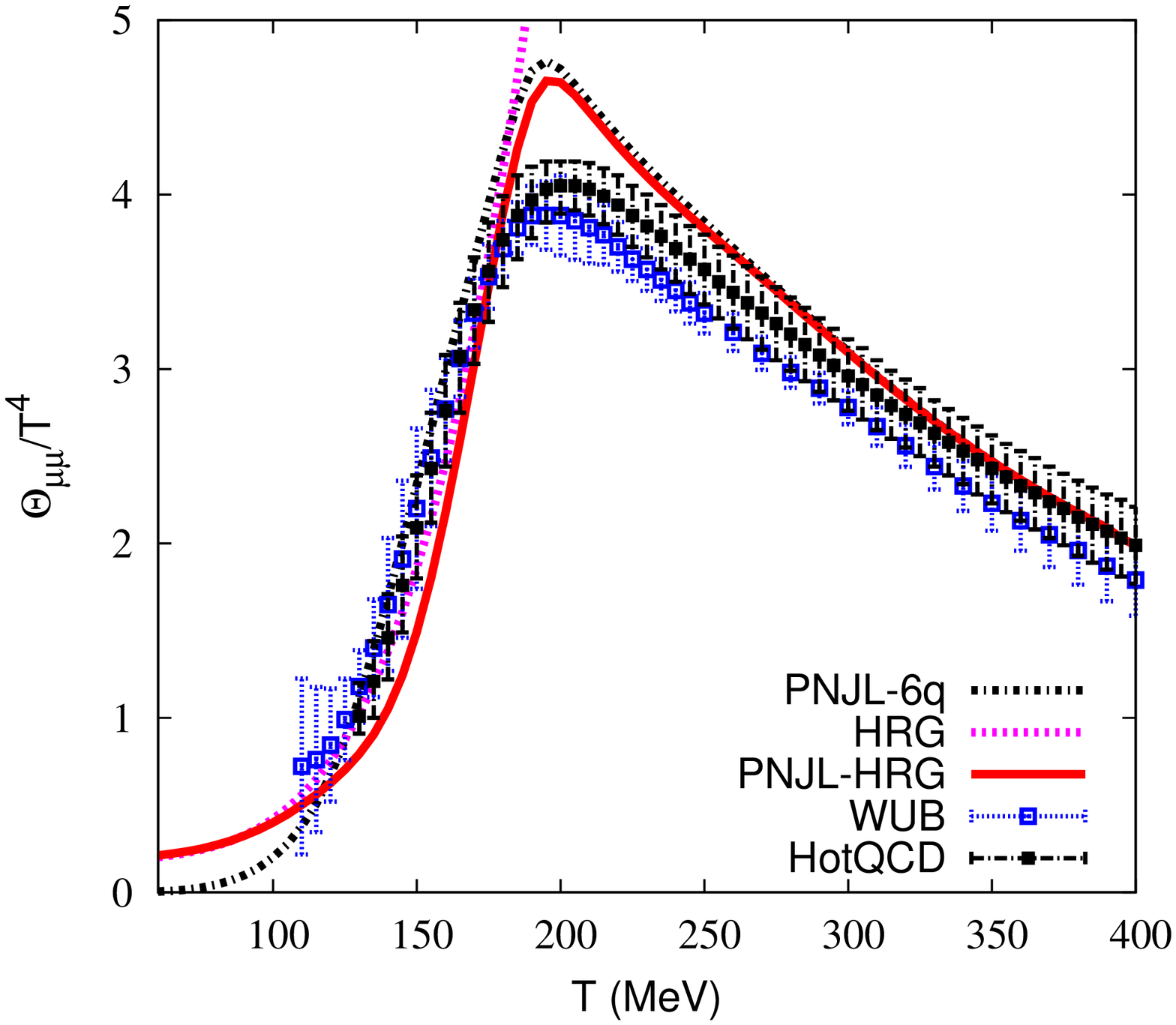}
\includegraphics[scale=0.28,angle=270]{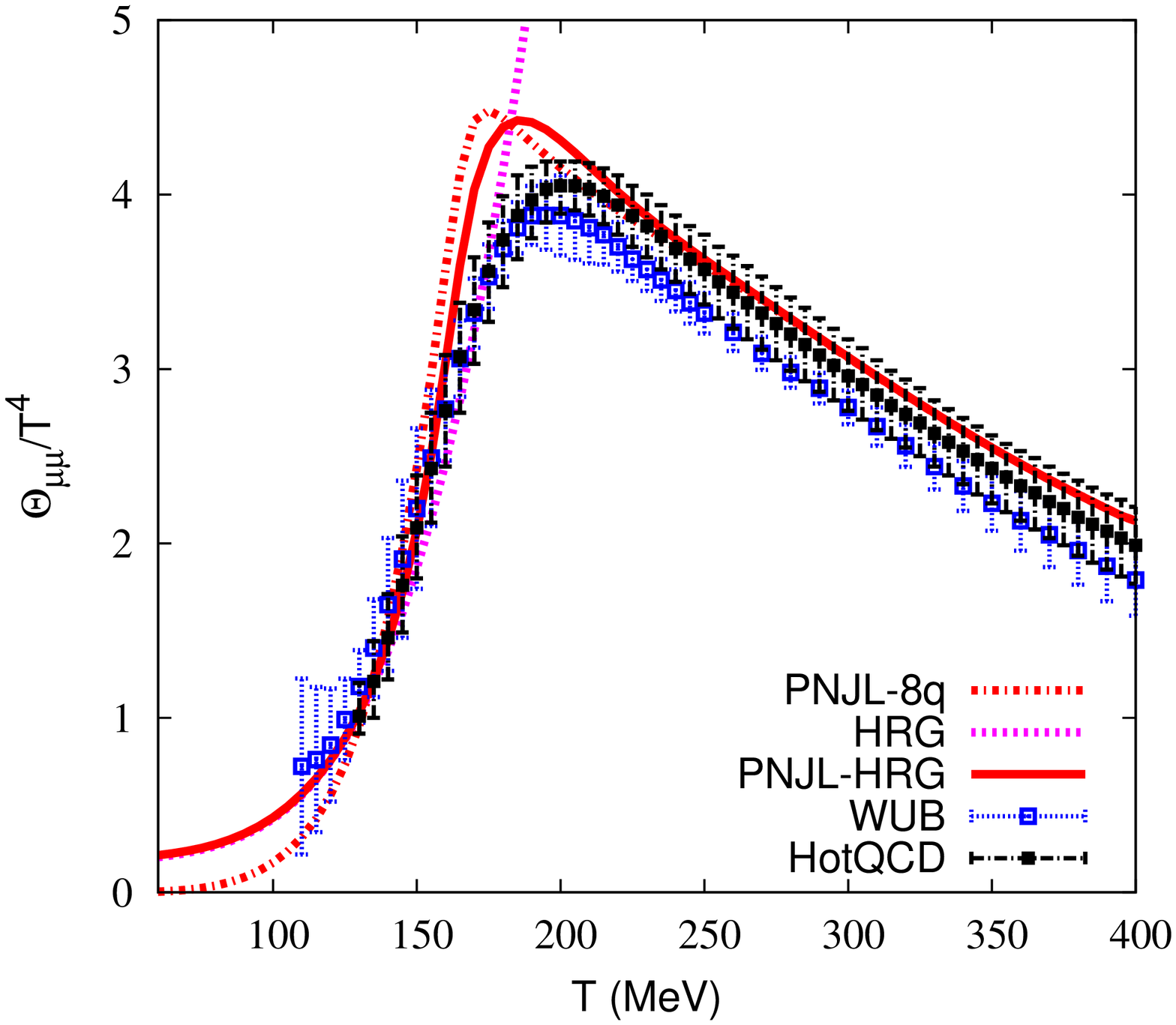}
\caption{(Color online) The scaled trace of energy-momentum tensor are
plotted as function of temperature. The upper panels are drawn without
derivatives of switching function and lower panels include those
derivatives. The left panel is for PNJL model with six quark
interactions and the right panel for PNJL model with eight quark
interactions. The continuum extrapolated lattice QCD data are taken from
Ref.~\cite{Bazavov14} (HotQCD) and Ref.~\cite{Borsanyi14} (WUB).}
\label{fig.trace}
\end{figure}

\begin{figure}[!htb]
\centering
\includegraphics[scale=0.28,angle=270]{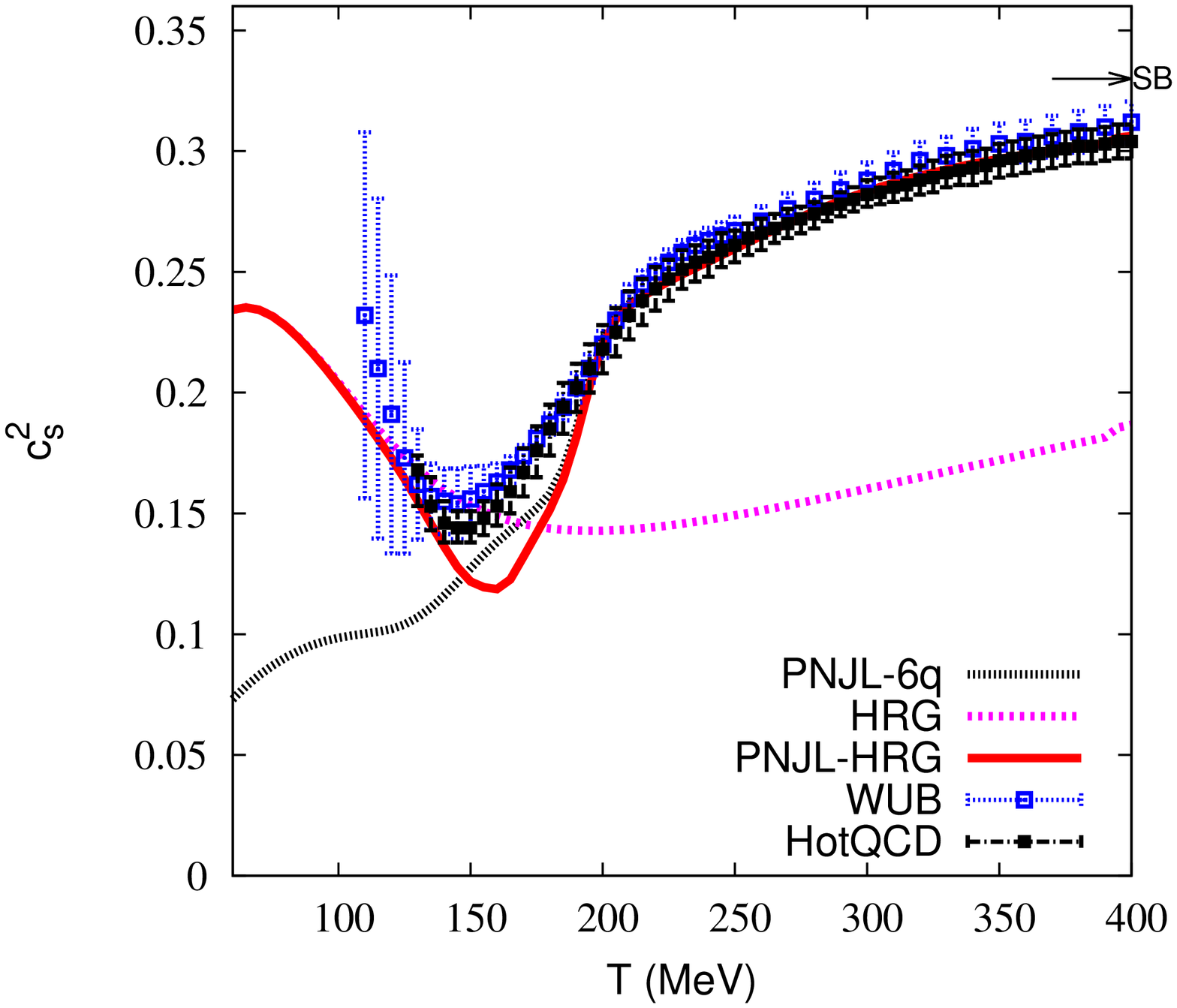}
\includegraphics[scale=0.28,angle=270]{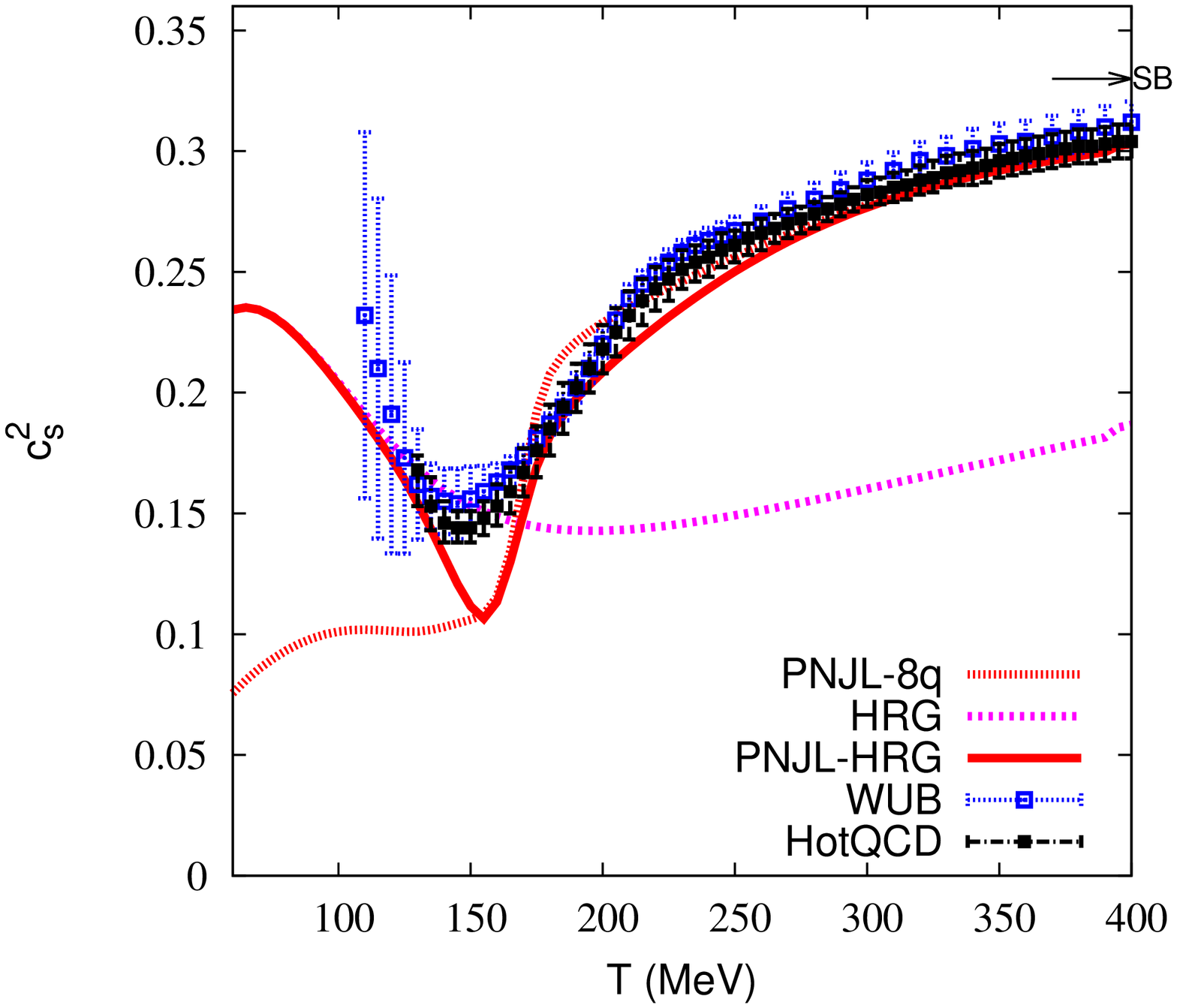}\\
\includegraphics[scale=0.28,angle=270]{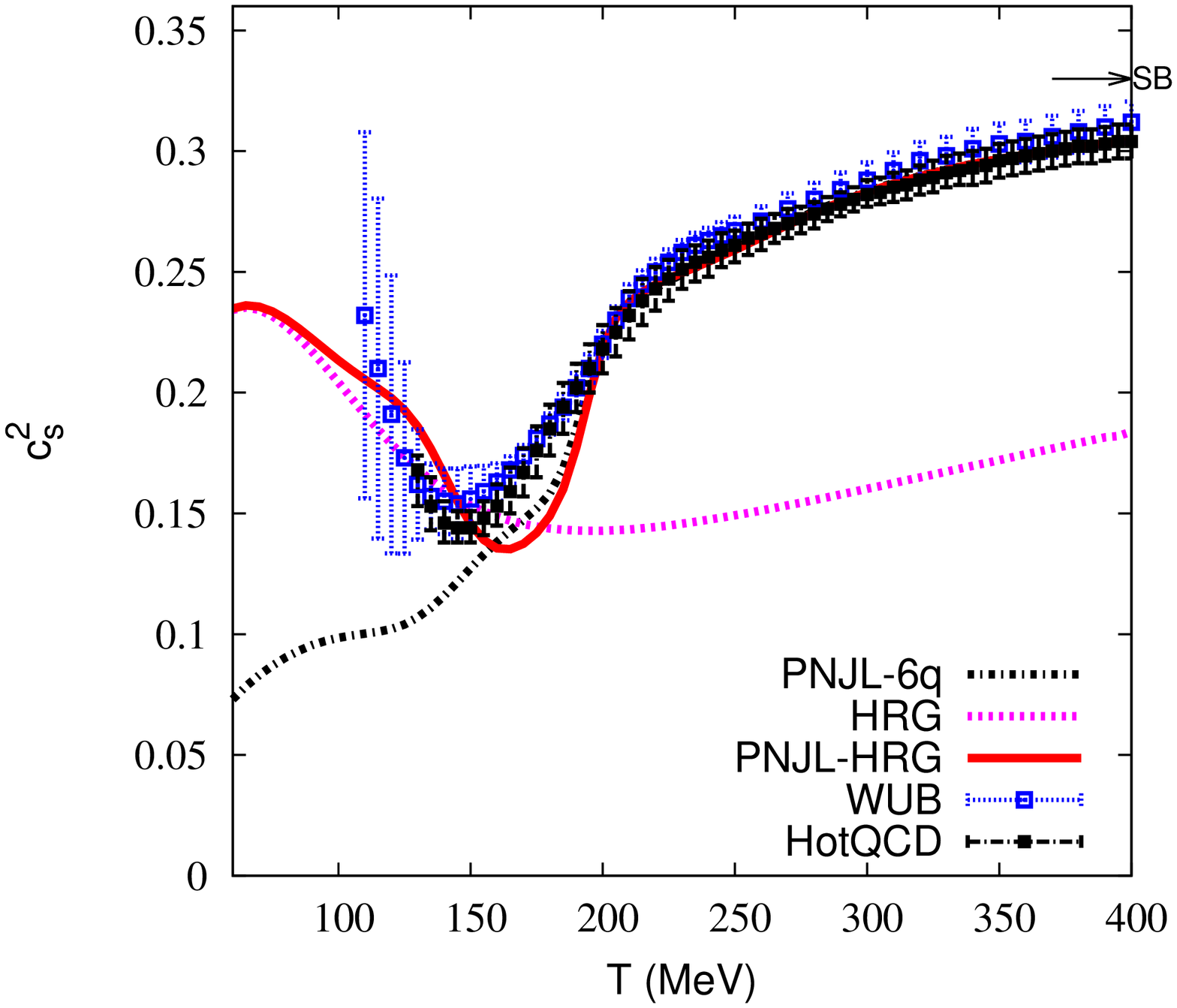}
\includegraphics[scale=0.28,angle=270]{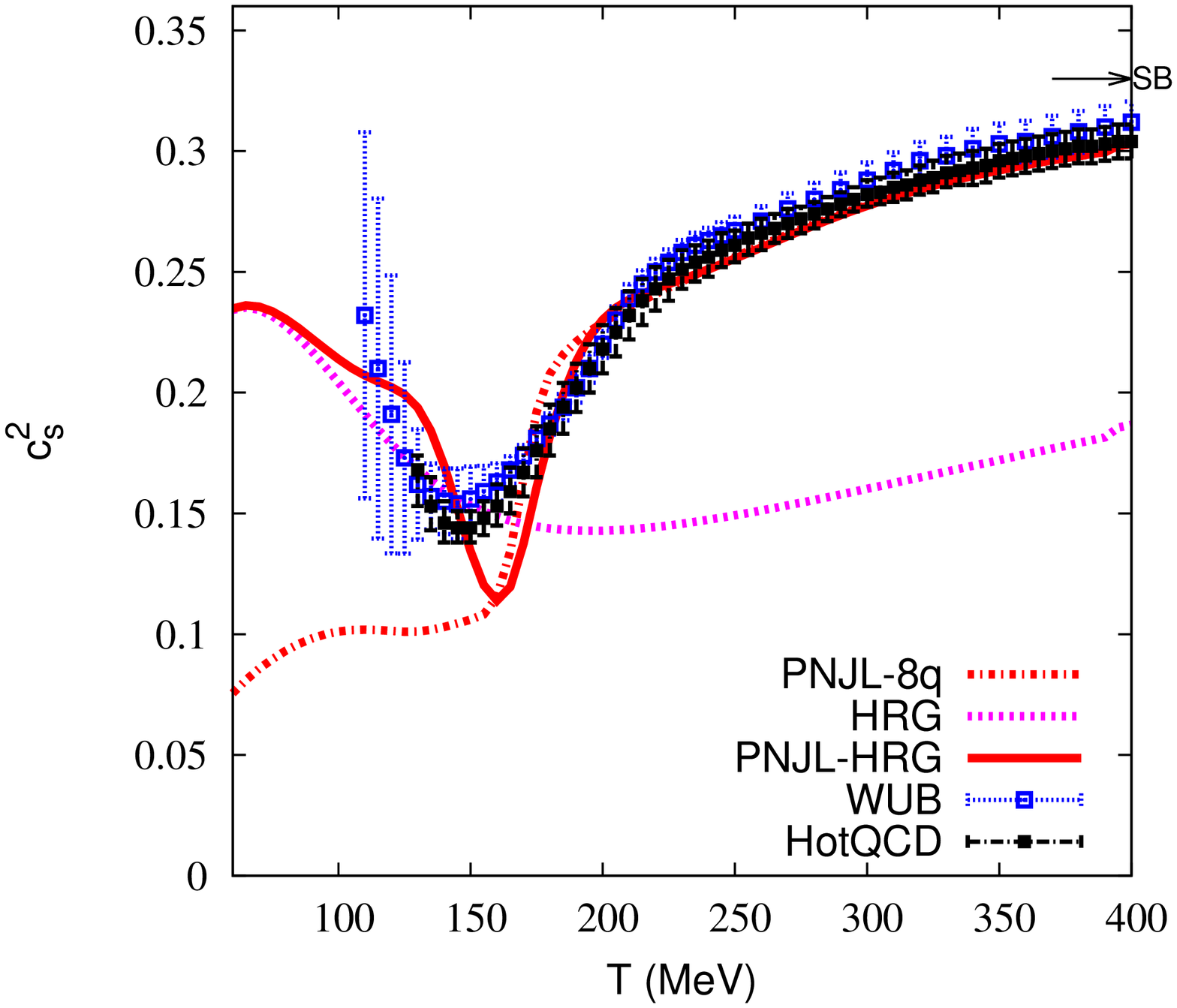}
\caption{(Color online) The squared speed of sound is plotted as a
function of temperature. The upper panels are drawn without derivatives
of switching function and lower panels include those derivatives.  The
left panel is for PNJL model with six quark interactions and the right
panel for PNJL model with eight quark interactions. The continuum
extrapolated lattice QCD data are taken from Ref.~\cite{Bazavov14}
(HotQCD) and Ref.~\cite{Borsanyi14} (WUB).}
\label{fg.cs2}
\end{figure}

\par
The two other quantities that we considered for critically assessing the
quantitative agreement with lattice QCD data are the trace of the
energy-momentum tensor and the squared speed of sound. The trace of the
energy-momentum tensor is defined as $\Theta_{\mu \mu}=(\epsilon-3P)$.
It is expected to be zero in a conformal theory. However the
introduction of any scale like the cross-over or transition temperatures
in the theory due to quantum interactions would break conformality and
the trace will be non-zero. This is expected to be a very sensitive
measure, inherent to the theory itself. Therefore a satisfactory
agreement of the results obtained from lattice QCD with any other model
is imperative for validation of the model as a simpler version of the
theory.

\par
In Fig.~(\ref{fig.trace}) we show scaled $\Theta_{\mu \mu}$ as a
function of temperature for various models and compare them to the
lattice QCD data ~\cite{Bazavov14, Borsanyi14}. Here the HRG model agrees
with lattice QCD data up to about $T=170$ MeV and rises much faster for
higher temperatures. The PNJL model agrees with lattice QCD data above
$T=150$ MeV, except near the peak at around $T=200$ MeV. Apart from this
peak region, the combined PNJL-HRG model now agrees with the lattice QCD
data for the full temperature range. The differences in the results due
to the type of PNJL model chosen is again quite small. However near the
peak region the eight-quark version of the PNJL model better describes
the lattice QCD data and so does the PNJL-HRG model. The deviation from
the lattice QCD data near the peak region would need further improvement
of the PNJL model as the switching function favors it over the HRG model
in this temperature range. Since our motivation here is to improve the
situation for the temperature region dominated by hadrons, we do not
intend to address this discrepancy here.

\par
We now discuss the behavior of the speed of sound as a function of
temperature. The squared speed of sound is given by,

\begin{equation}
c_s^2 =\frac{\partial P}{\partial \epsilon} = \frac{s}{C_V} 
\label{eq.cs2}
\end{equation}

\noindent
The variation of the squared speed of sound with temperature is shown in
Fig.~(\ref{fg.cs2}) for various models and compared with the lattice QCD
data ~\cite{Bazavov14, Borsanyi14}. As seen from Eq.~\ref{eq.cs2},
the speed of sound on one hand contains the information of the equation
of state and on the other hand contains the information of entropy to
specific heat ratio. Thus the characteristics of the speed of sound
depends strongly on the phase of strongly interacting matter. One
expects that at very low temperatures the speed of sound would be small
as the pressure of the hadronic system with masses much larger than the
system is negligible. With increase in temperature the speed of sound
will increase. However with increasing temperature the hadron resonances
with higher and higher masses would be excited and the speed of sound
may not reach the SB limit. In fact it may even start decreasing with
temperature \cite{Andronic:2012ut}. Using the HRG model all these
features are obtained as shown in Fig.~\ref{fg.cs2}. After the
crossover, the degrees of freedom are supposed to change from hadronic
to partonic ones and therefore speed of sound may again increase. The
minimum of the speed of sound known as the softest point may be a
crucial indicator of the transition to be observed in heavy-ion
collisions \cite{Hung95}.  Such a minimum in the temperature variation
of speed of sound is visible in the lattice QCD data as shown in
Fig.~\ref{fg.cs2}, but is clearly absent in the PNJL model results. It
is interesting to see that even the HRG model has a soft point of the
same order of magnitude but at a higher temperature compared to the
lattice QCD data.  While the HRG model agrees with lattice QCD data up
to about $T=150$ MeV, the PNJL model results are consistent with the
lattice data above $T_c$. Thus the combined PNJL-HRG model shows
reasonable agreement with the lattice QCD data in the full range of
temperatures.

\par
From the above discussions we may conclude that a hybrid model like the
PNJL-HRG satisfactorily describes  the equation of state
of strongly interacting matter in a wide range of temperatures. We would
now like to investigate whether the same holds true for the various
fluctuations and correlations at vanishing chemical potentials.

\section{Fluctuations and correlations of conserved charges}
\label{sc.fluccor}

\begin{figure}[!htb]
\includegraphics[scale=0.29,angle=270]{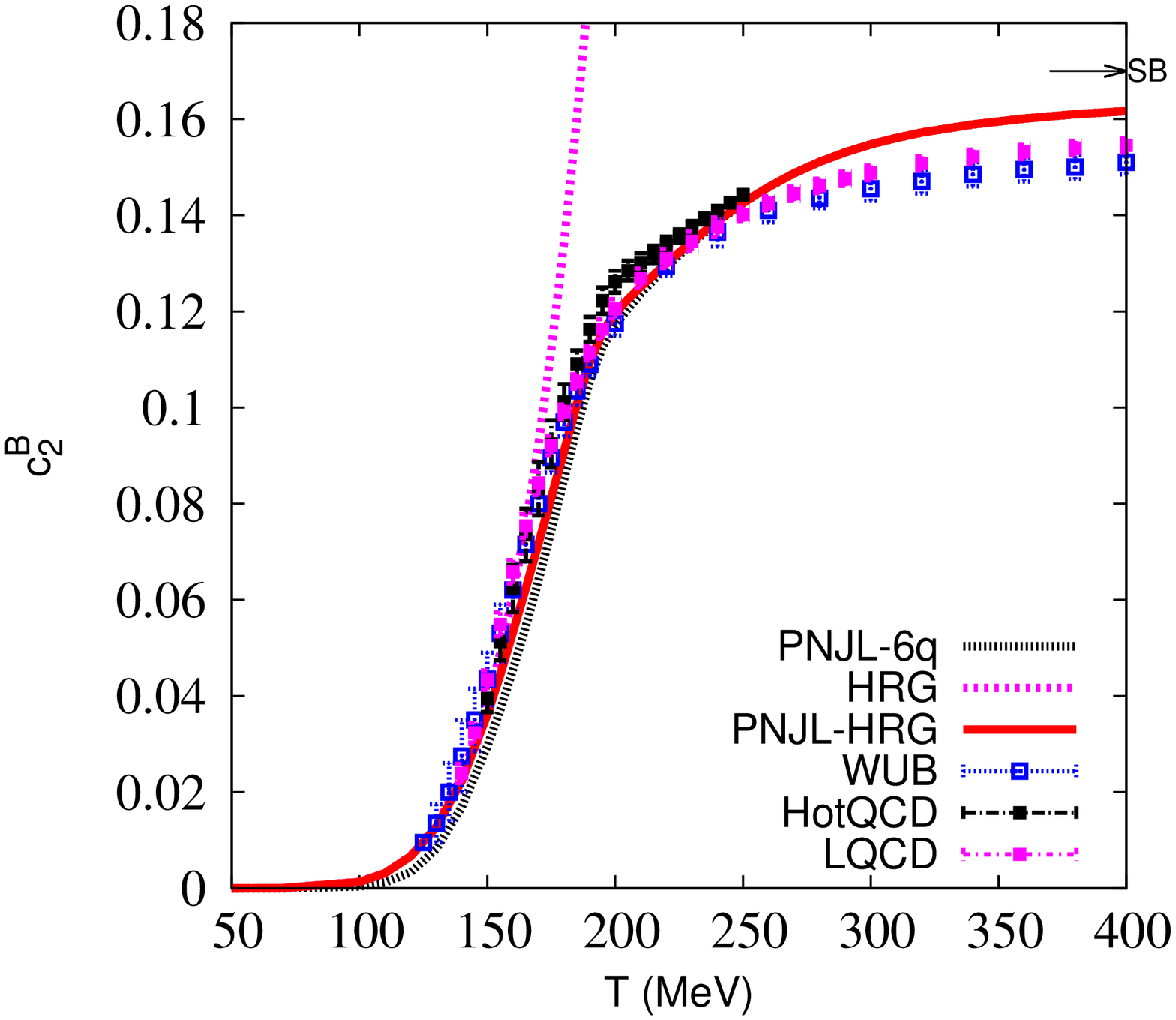}
\includegraphics[scale=0.29,angle=270]{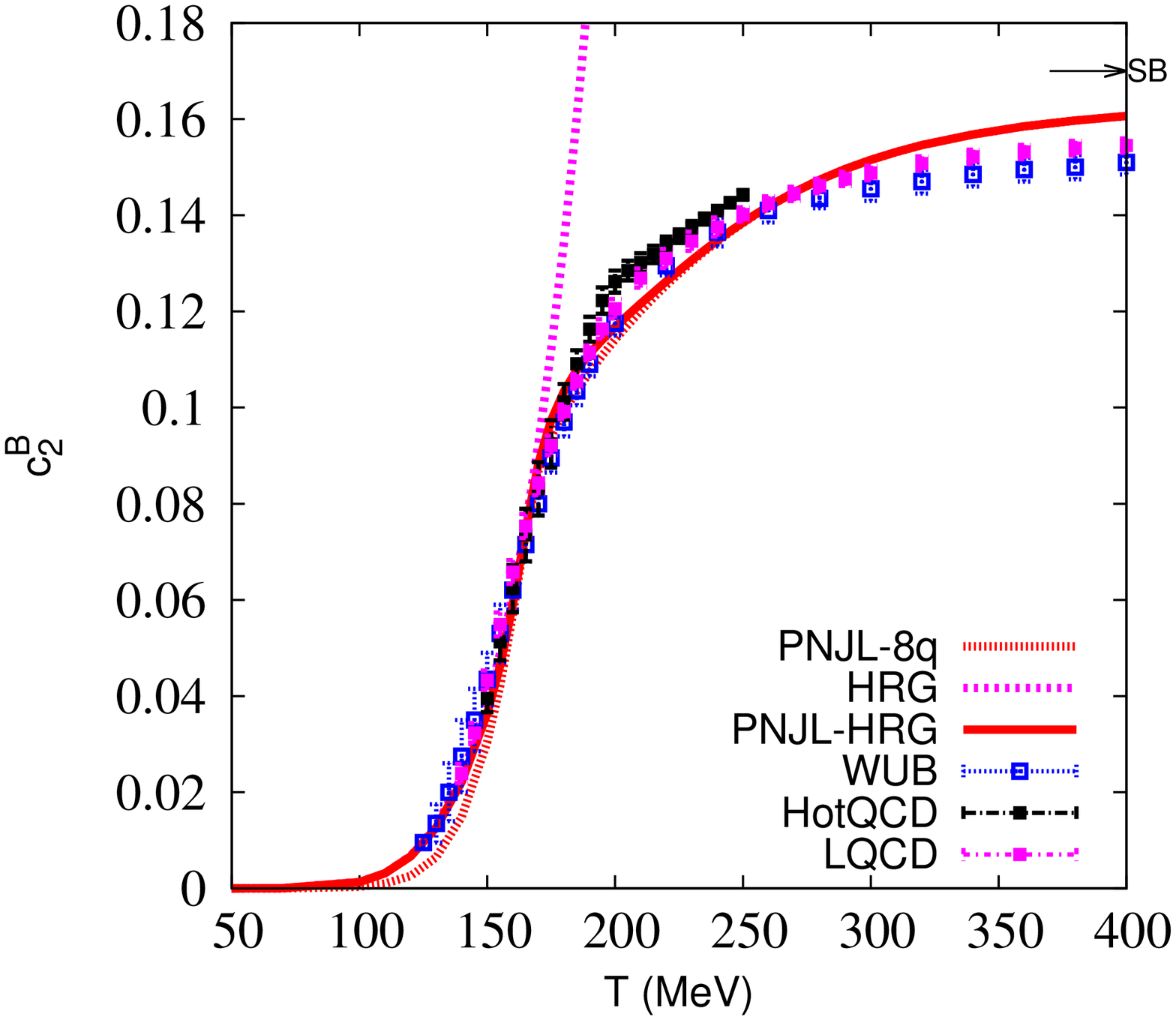}
\caption{(Color online) Baryon number fluctuation as function of
temperature. The continuum extrapolated lattice data are from
Ref.~\cite{Bazavov:2012jq} (HotQCD), Ref.~\cite{Borsanyi:2011sw} (WUB)
and Ref.~\cite{Bellwied2} (LQCD).}
\label{fg.cB}
\end{figure}

\par
In strong interactions the net number of various quark flavors are
conserved. But since quark states are not physically observed, one has
to relate these flavor conservations into conservations of the
experimentally observed charges of hadrons like baryon number $B$,
electric charge $Q$ and strangeness $S$. The fluctuations and
correlations of conserved charges depend significantly on the state of
strongly interacting matter at high temperatures and densities
~\cite{Koch:2008ia, Deb, Lahiri, Abelev, Ejiri}.  Variations of these
quantities with temperature and various chemical potentials are supposed
to carry signatures of phase transition or crossover ~\cite{Shuryak,
Jeon, Jeon1, Heinz, Kinkar, Bhattacharyya:2013oya, Sur, Bhatta,
Bhattacharyya:2015zka, Upadhaya, Hatta}. They are related to the
susceptibilities of pressure via fluctuation dissipation
theorem~\cite{Anirban}. The susceptibilities at various orders are
easily obtained as a Taylor series expansion of pressure in terms of the
corresponding chemical potentials. The diagonal Taylor coefficients
$c_n^X(T)$ ($X=B,Q,S$) of $n^{th}$ order for the scaled pressure
$P(T,\mu_B,\mu_Q,\mu_S)/T^4$ may be written in terms of the fluctuations
$\chi_n^X(T)$ of the corresponding order as,

\begin{equation}
c_n^X(T)=\frac{1}{n!}\frac{\partial^n(P/T^4)}
{\partial(\frac{\mu_X}{T})^n}=T^{n-4}\chi_n^X(T)
\end{equation}

\noindent
where the expansion is carried out around $\mu_B=\mu_Q=\mu_S=0$.
Similarly, the off-diagonal coefficients $c_{n,m}^{X,Y}(T)$
($X,Y=B,Q,S$; $X\ne Y$) of the $(m+n)^{th}$ order in the Taylor
expansion of scaled pressure are related to the correlations between the
conserved charges $\chi_{n,m}^{X,Y}(T)$ as,

\begin{equation}
c_{m,n}^{X,Y}=\frac{1}{m!n!}\frac{\partial^{m+n}(P/T^4)}
{(\partial(\frac{\mu_X}{T})^m)(\partial(\frac{\mu_Y}{T})^n)}
=T^{m+n-4}\chi_{n,m}^{X,Y}(T)
\end{equation}

\par
At zero chemical potentials some of these fluctuations and correlations
have been measured in the lattice QCD framework either in the continuum
limit ~\cite{Borsanyi:2011sw, Bazavov:2012jq, Katz1, Bellwied1, Swagato1,
Swagato2, Bellwied2} or for small lattice spacings close to the continuum
limit~\cite{Bazavov1213}. We shall now discuss how much these quantities
obtained in the PNJL-HRG model agree with the lattice QCD data.  In the
PNJL model the fluctuations and correlations are obtained by a suitable
Taylor series fitting as has been discussed in detail in~\cite{Ray}. On
the other hand these quantities are obtained easily in the HRG model by
taking the corresponding derivatives with respect to the chemical
potentials. Once the fluctuations and correlations are obtained in both
the models, they are combined into the PNJL-HRG model using the same
switching function $S(T)$ identified in last section. They are given by,

\begin{eqnarray}
c_n^X&=&S(T) {c_n^X}_P+(1-S(T)) {c_n^X}_H\\
{\rm and~~}
c_{m,n}^{X,Y}&=&S(T) {c_{m,n}^{X,Y}}_P+(1-S(T)) {c_{m,n}^{X,Y}}H
\end{eqnarray}

\noindent
As mentioned earlier we have assumed the switching function to be
independent of the chemical potential in the close vicinity of zero
chemical potentials. Let us now discuss some of these fluctuations and
correlations obtained in the models and compare them to the lattice QCD
data.

\par
In Fig.~\ref{fg.cB} the variation of the baryon number susceptibility
$c_2^B$ is shown as a function of temperature for various models and
compared with lattice QCD data. Here we find $c_2^B$ obtained in HRG
model agrees with lattice QCD data up to about $T\sim170$ MeV. The PNJL
model also agrees with lattice QCD quite well in this temperature range
and therefore also the PNJL-HRG model. At higher temperatures the
hybrid model is dominated by the PNJL model dynamics. There is a slight
overestimation in the PNJL model which is also reflected in the PNJL-HRG
model. Apart from that, the results in the six-quark and eight-quark
versions of the PNJL models are numerically commensurate. The baryonic
contribution of the PNJL model therefore seems to be sufficient in
describing the strongly interacting matter even in the low temperature
region described either by HRG model or the lattice QCD data.

\begin{figure}[!htb]
\includegraphics[scale=0.29,angle=270]{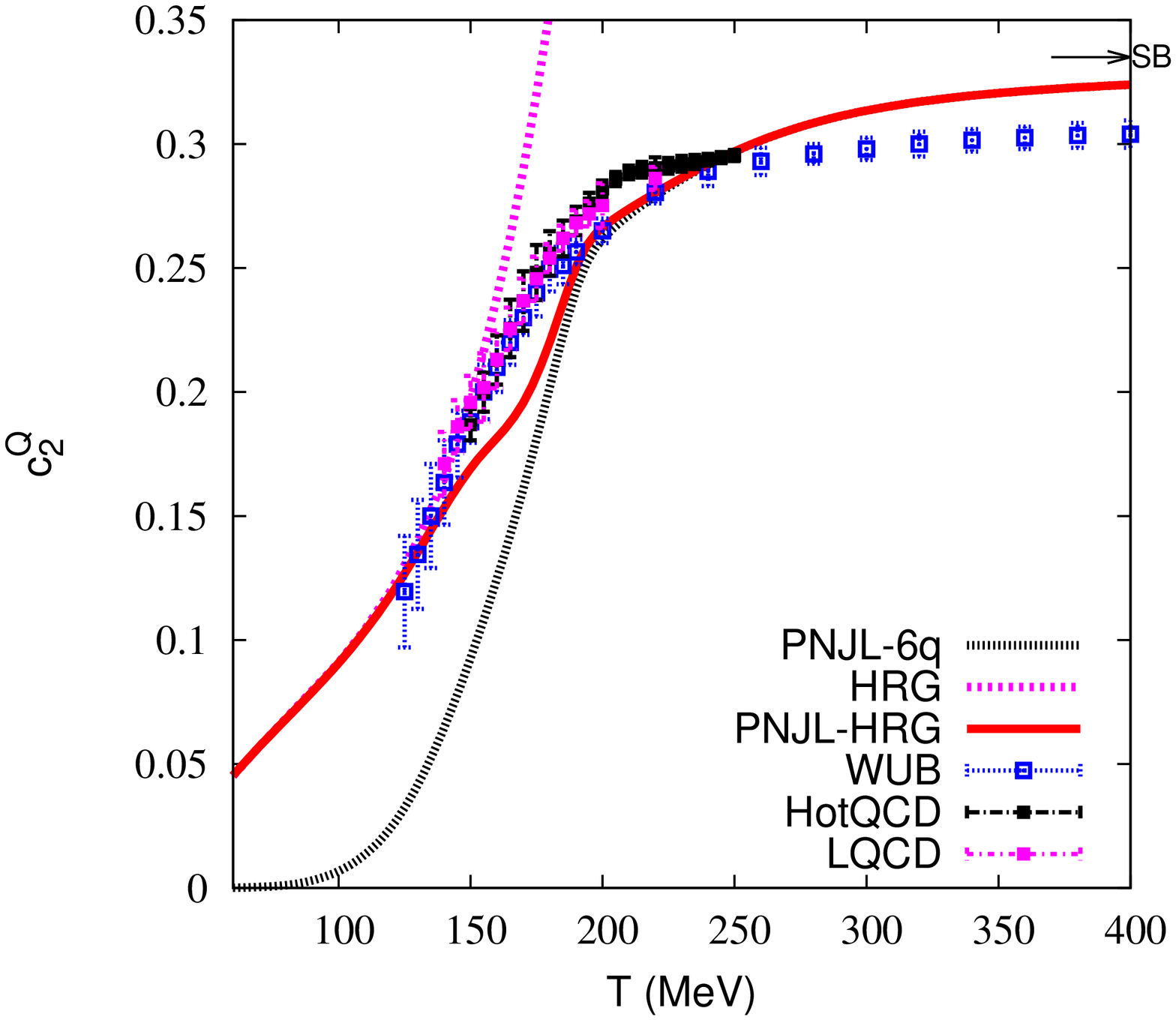}
\includegraphics[scale=0.29,angle=270]{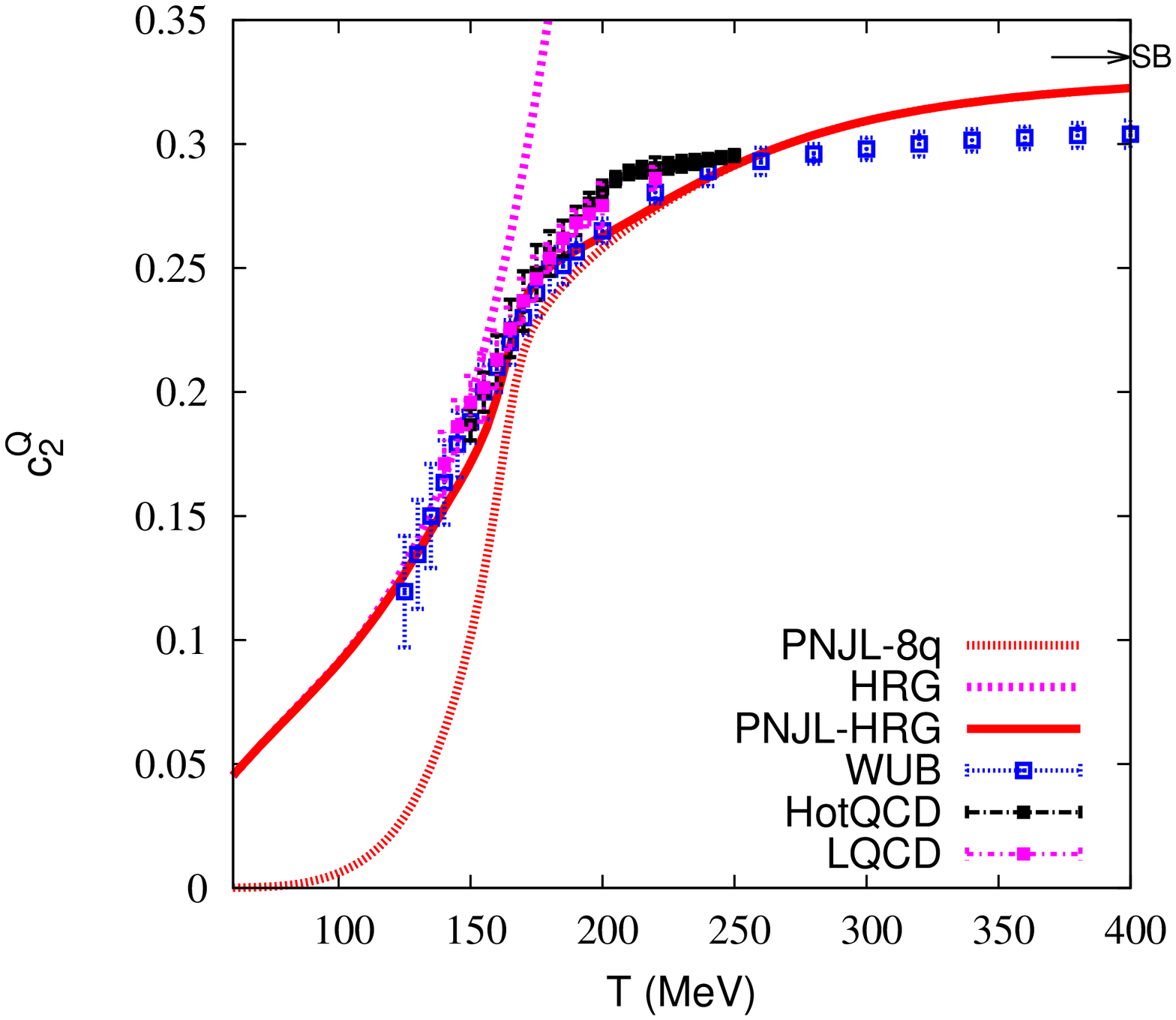}
\caption{(Color online) Electric charge fluctuation as function of
temperature. The continuum extrapolated lattice data are from
Ref.~\cite{Bazavov:2012jq} (HotQCD), Ref.~\cite{Borsanyi:2011sw} (WUB)
and \cite{Bellwied2} (LQCD).}
\label{fg.cQ}
\end{figure}

\begin{figure}[!htb]
\includegraphics[scale=0.29,angle=270]{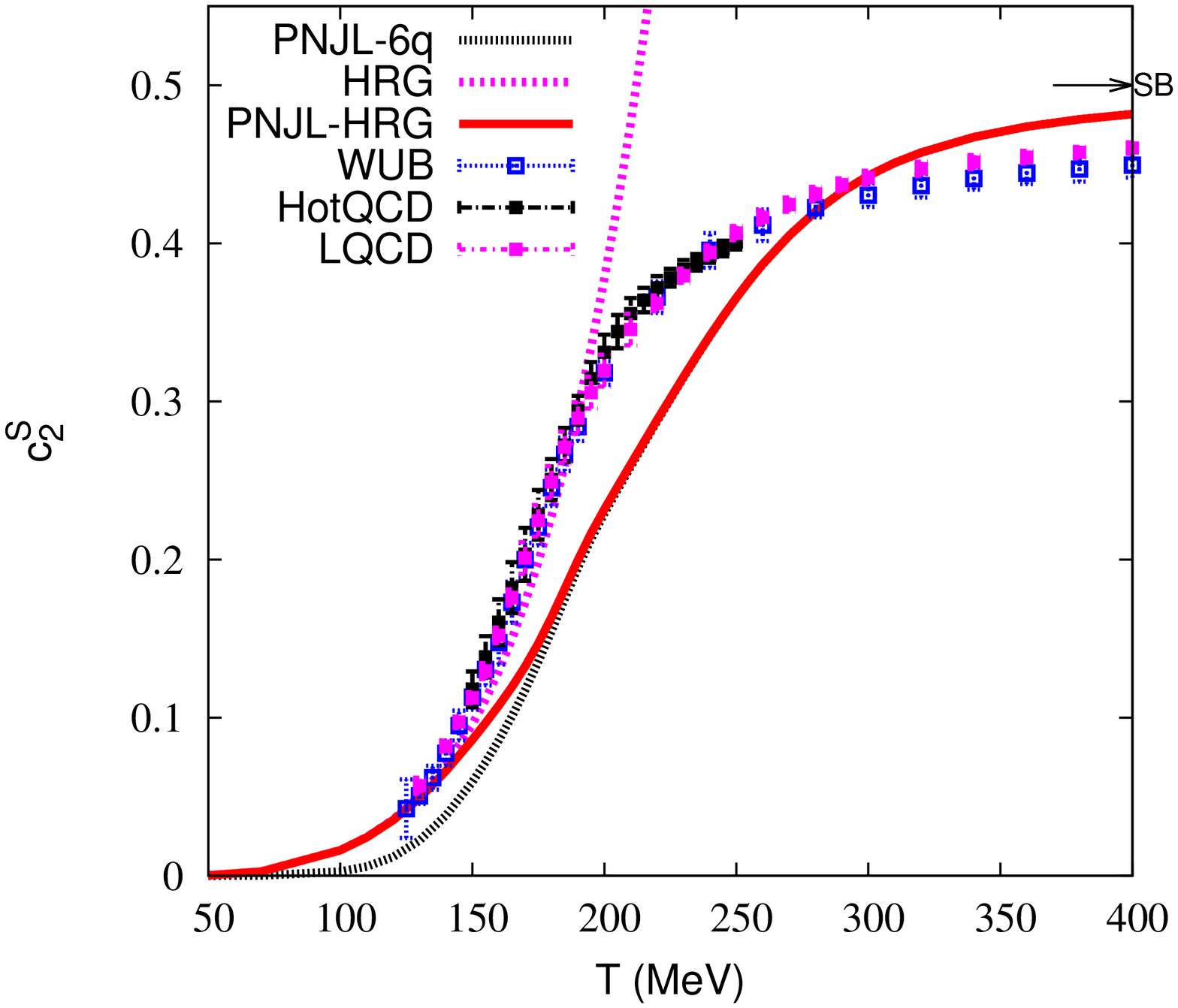}
\includegraphics[scale=0.29,angle=270]{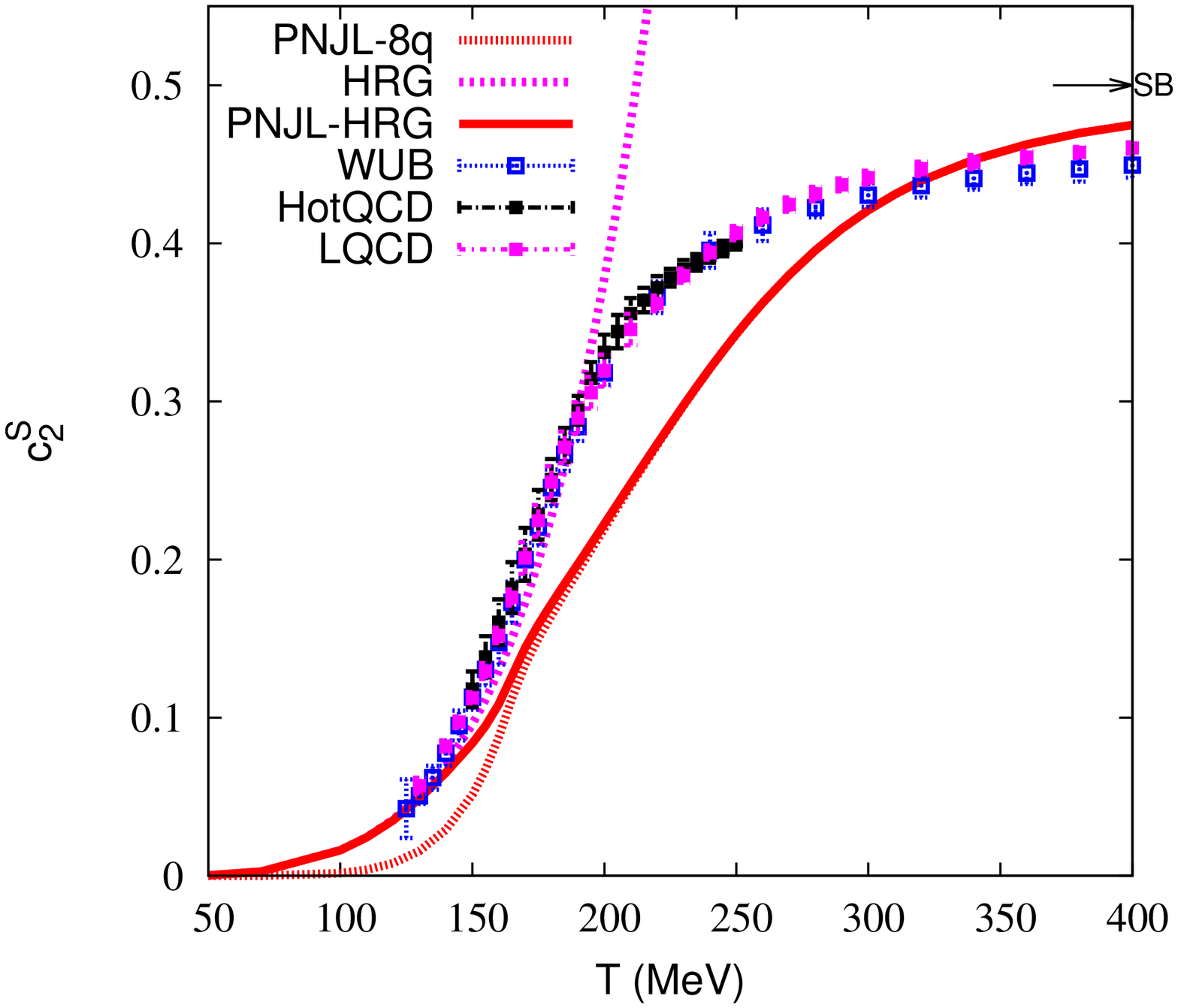}
\caption{(Color online) Strangeness fluctuation as function of
temperature. The continuum extrapolated lattice data are from
Ref.~\cite{Bazavov:2012jq} (HotQCD), Ref.~\cite{Borsanyi:2011sw} (WUB)
and Ref.~\cite{Bellwied2} (LQCD).}
\label{fg.cS}
\end{figure}

\par
The variation of the electric charge susceptibility with temperature is
shown in Fig.~\ref{fg.cQ}. There is a significant difference between
PNJL model and lattice QCD results for $c_2^Q$ below the crossover
temperature $T_c$. The lattice data is much larger than the PNJL model
results. Though the baryon fluctuations in the lattice data are well
accounted for by the constituent quarks in the PNJL model, proper
considerations of other hadronic degrees of freedom below $T_c$ is
crucial to obtain the correct values of electric charge fluctuations.
On the other hand lattice QCD data is reproduced by the HRG model very
well for $T<150$ MeV. This is expected as the charge sector has dominant
contributors from the light hadrons, which are practically absent in the
PNJL model. Once the PNJL and HRG models are combined using the
switching function, the PNJL-HRG model results again agree with the
lattice QCD data very well.

\par
The temperature variation of the strangeness susceptibility $c_2^S$ is
shown in Fig.~\ref{fg.cS}. The computations in the HRG model seem to
agree with the lattice QCD data up to much higher temperatures
$T\sim190$ MeV.  This is surprising given that the crossover
temperature, as well as the temperature around which almost all other
quantities computed in the HRG model start deviating from lattice QCD
data at around $T\sim160$ MeV. One would expect the $c_2^S$ in the HRG
model to rise much faster and start deviating from lattice QCD data at
much lower temperatures.  One of the possible reasons for such a result
may be that HRG model is constructed from experimentally observed
hadrons, whereas the lattice QCD formulation could have contributions
from additional species of strange hadrons, which are also predicted by
quark model calculations~\cite{Bazavov14a}.

\par
At the same time the quantitative results for $c_2^S$ obtained in PNJL
model are significantly different from lattice QCD data up to $T\sim250$
MeV. The constituent masses of the strange quarks in the PNJL model have
values above $500$ MeV for $T<150$ MeV. This is consistent with the
omega baryon mass. However as the temperature rises, the constituent
masses of strange quarks do not fall as fast as that of the light
quarks. Therefore the strangeness fluctuations on their part, do not
rise as fast as that of the light quarks as observed in the lattice QCD
data. This aspect of the PNJL model would need further scrutiny and will
be discussed elsewhere.  Note that the range of temperature where the
PNJL model disagrees with lattice QCD data is above the range where the
switching from HRG model to PNJL model takes place. Inclusion of HRG
model is therefore insufficient to address this issue.  As a result the
hybrid PNJL-HRG model does not agree well with the lattice QCD data.  We
thus encounter the first observable where the PNJL-HRG model could not
satisfactorily describe the lattice QCD data in the full temperature
range. It is apparent that this may be the case for other quantities
that are strongly dependent on the strangeness content of the PNJL
model.

\begin{figure}[!htb]
\includegraphics[scale=0.23,angle=270]{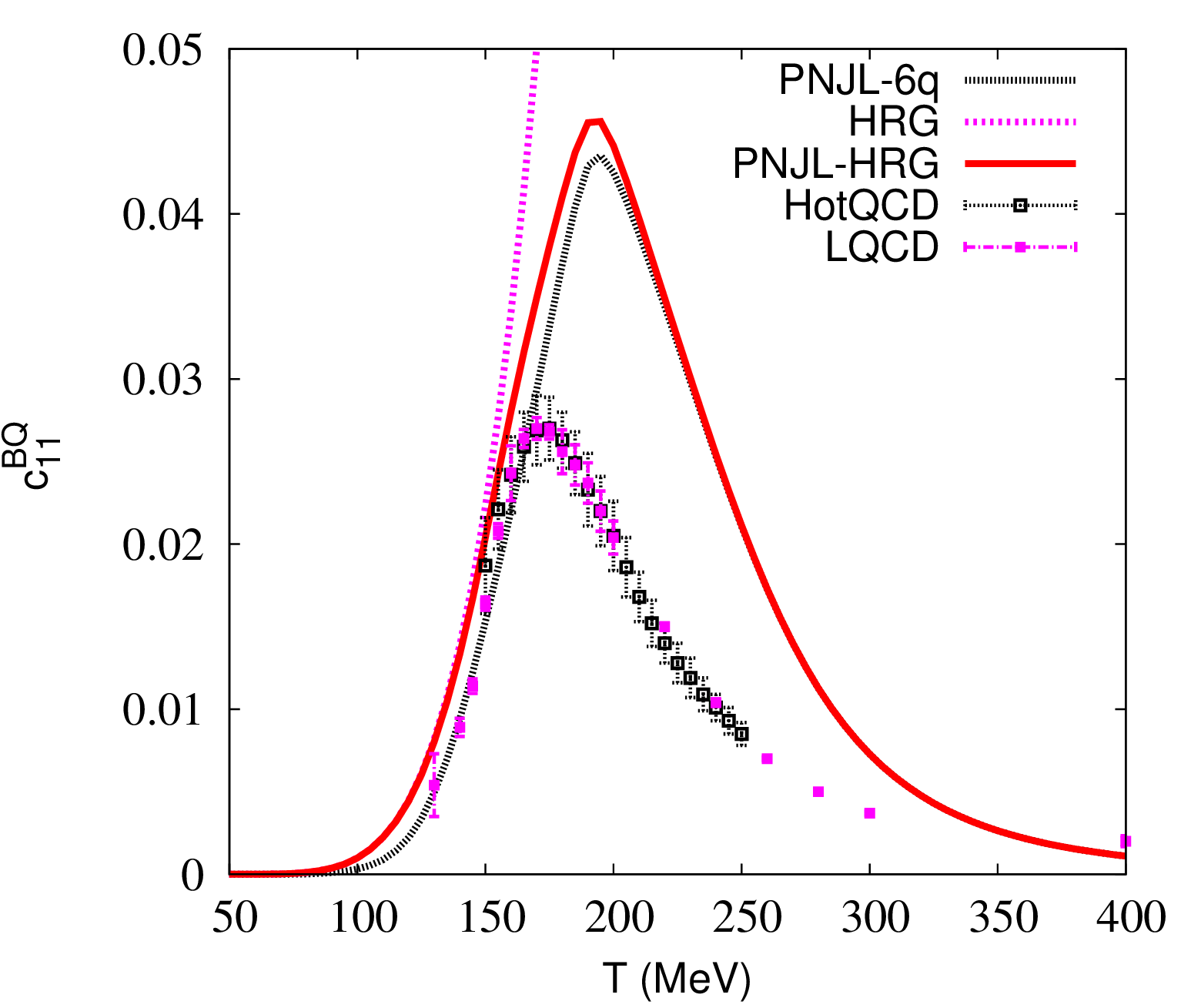}
\includegraphics[scale=0.23,angle=270]{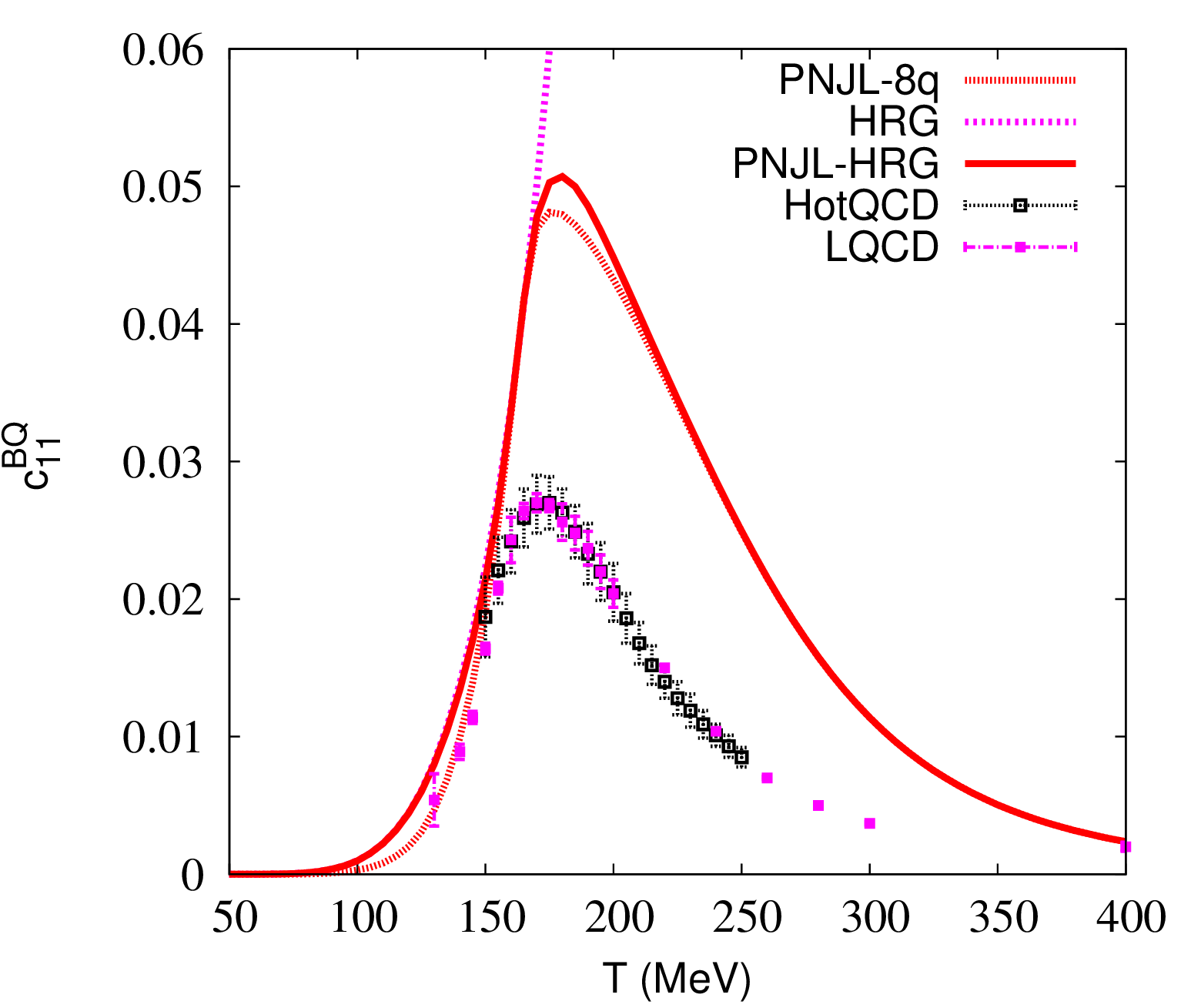}
\caption{(Color online) Baryon-electric charge correlation as function
of temperature. The continuum extrapolated lattice data are from
Ref.~\cite{Bazavov:2012jq} (HotQCD) and Ref.~\cite{Bellwied2} (LQCD).}
\label{fg.corrbq}
\end{figure}

\begin{figure}[!htb]
\includegraphics[scale=0.23,angle=270]{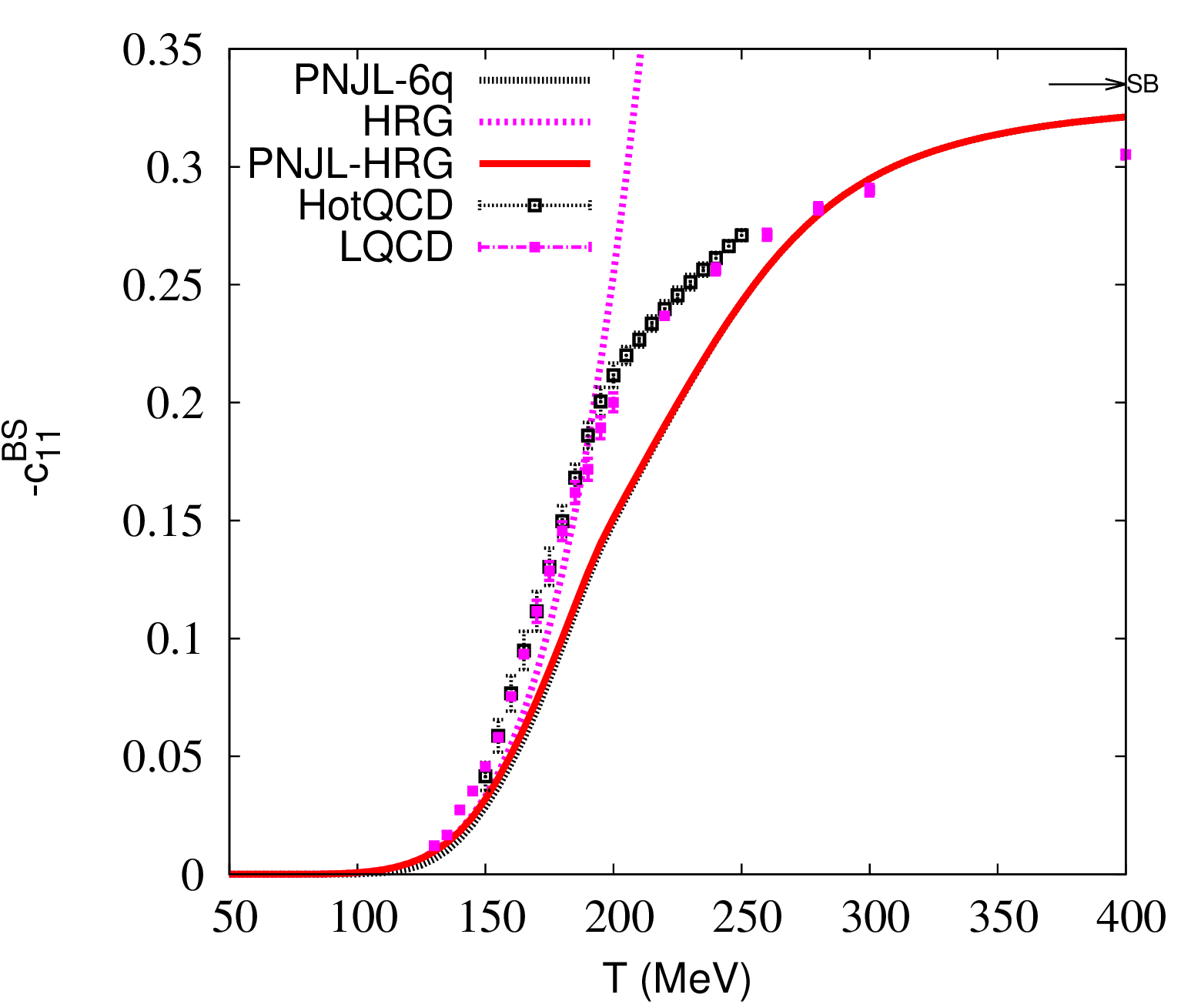}
\includegraphics[scale=0.23,angle=270]{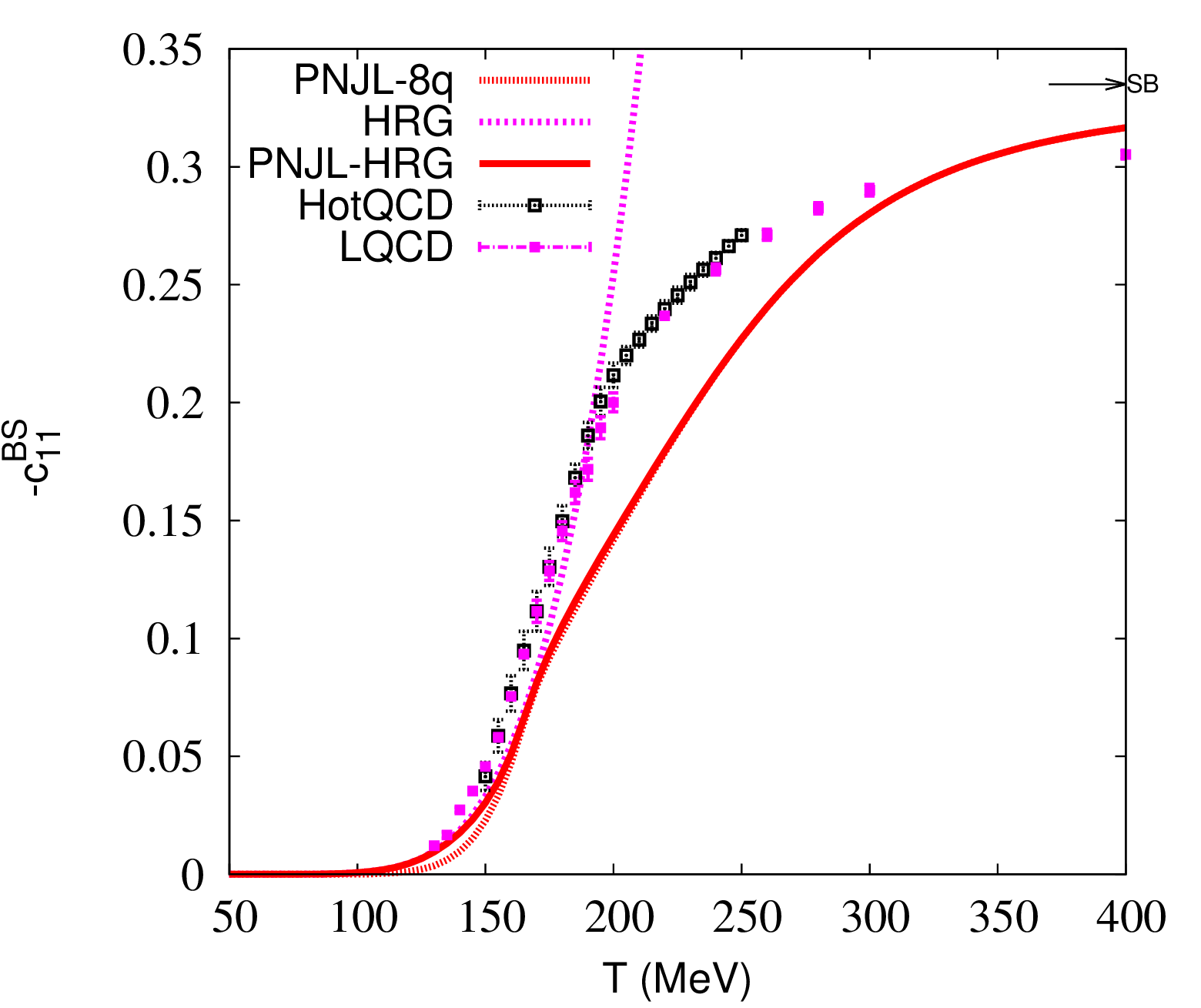}
\caption{(Color online) Baryon-strangeness correlation as function
of temperature. The continuum extrapolated lattice data are from
Ref.~\cite{Bazavov:2012jq} (HotQCD) and Ref.~\cite{Bellwied2} (LQCD).}
\label{fg.corrbs}
\end{figure}

\begin{figure}[!htb]
\includegraphics[scale=0.23,angle=270]{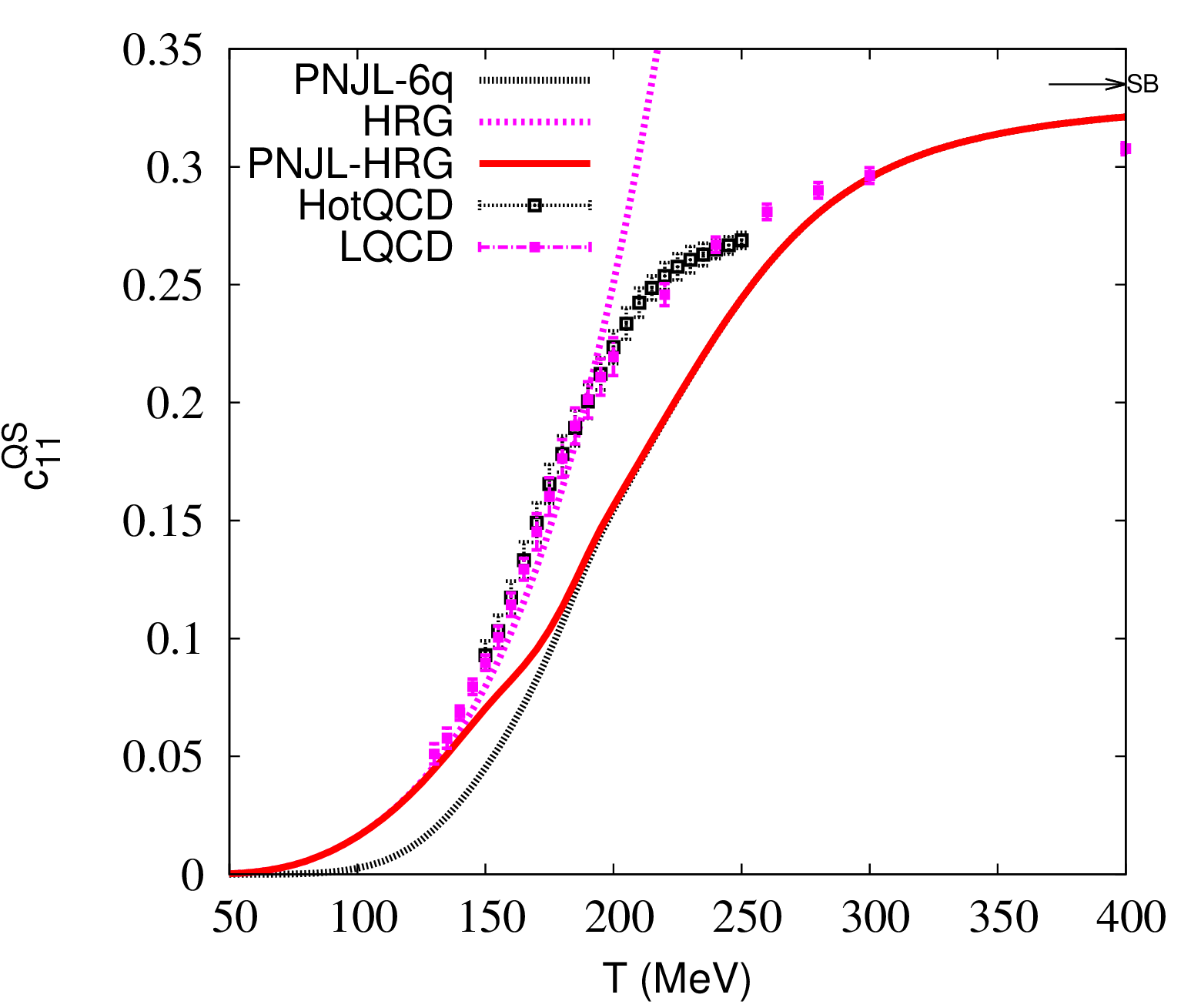}
\includegraphics[scale=0.23,angle=270]{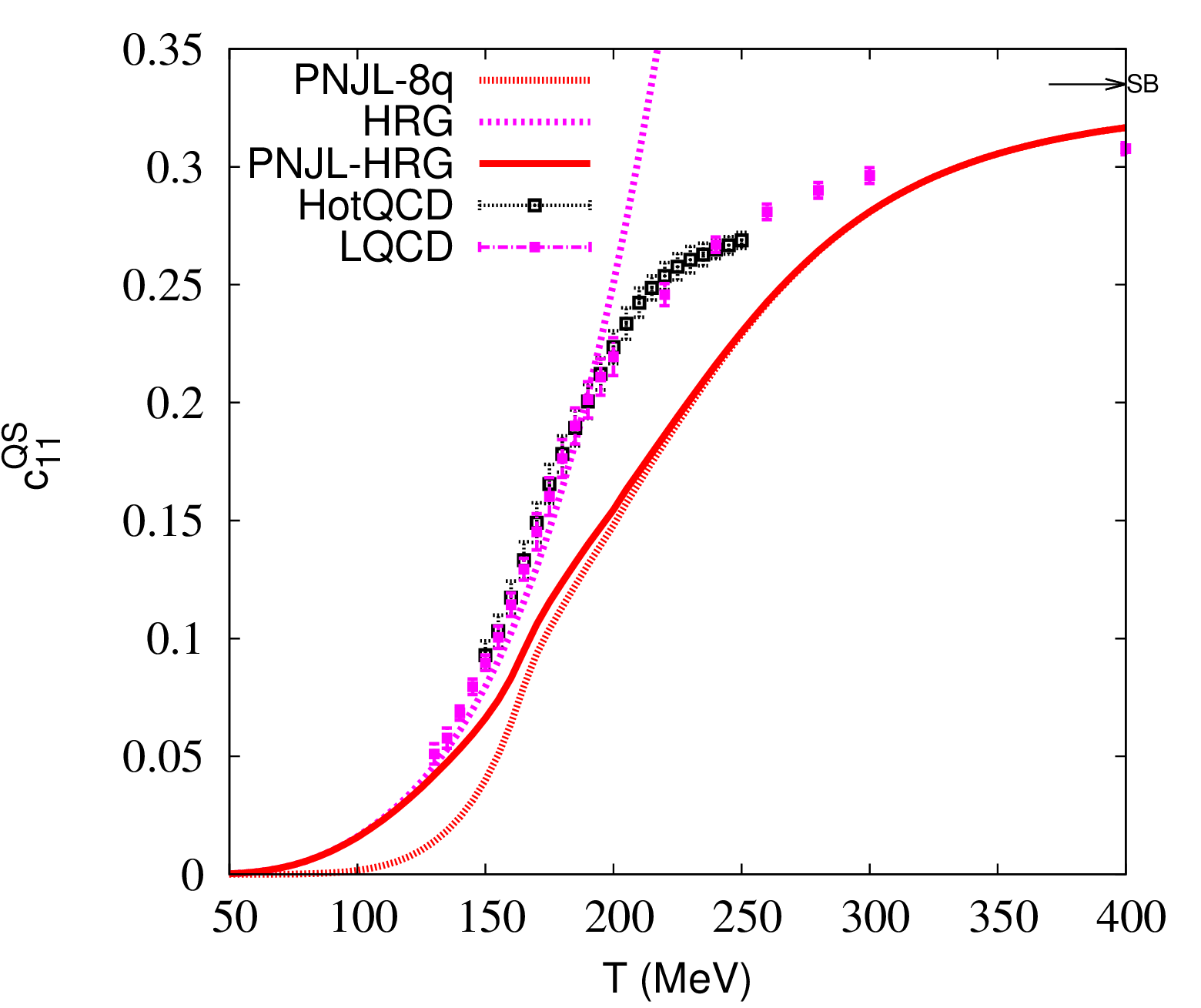}
\caption{(Color online) Electric charge-strangeness correlation as
function of temperature. The continuum extrapolated lattice data are
from Ref.~\cite{Bazavov:2012jq} (HotQCD) and Ref.~\cite{Bellwied2}
(LQCD).}
\label{fg.corrqs}
\end{figure}

\par
We now discuss the leading order correlations between the conserved
charges.  The correlator $c_{11}^{BQ}$ between the baryon number and
electric charge is shown in Fig.~\ref{fg.corrbq}.  In the hadronic phase
the baryon and electric charges remain correlated as the baryons have
positive electric charge and anti-baryons have negative electric charge.
For small temperatures the correlations are small due to the relatively
large masses. With increasing temperature however the correlation
becomes non-zero. On the other hand for the 2+1 flavor system there are
three quarks with equal baryon number in the partonic phase, but
electric charge of $down$ and $strange$ quarks are together opposite of
that of the $up$ quark. At large temperatures when the quark masses are
small with respect to the temperature, the BQ correlation should again
tend to zero. In HRG model the BQ correlations keep on increasing as
higher and higher mass states are getting excited with increase in
temperature. On the other hand for lattice QCD data and PNJL model as
well as any other model having a transition from hadronic to partonic
phases, $c_{11}^{BQ}$ would show a hump around the crossover region.
This is shown in Fig.~\ref{fg.corrbq}. Here the BQ correlation in the
PNJL model is larger than that obtained in the lattice QCD data for
$T>150$ MeV. It may again be anticipated that this is due to the slow
decrease of the strange quark mass with temperature in the PNJL model.
The number of strange quarks is therefore much smaller than that of the
light quarks, and they cannot compensate the electric charge of the up
quarks sufficiently.

\par
The baryon number to strangeness (BS) correlation $c_{11}^{BS}$ is shown
in Fig.~(\ref{fg.corrbs}) and the electric charge to strangeness (QS)
correlation $c_{11}^{QS}$ is shown in Fig.~(\ref{fg.corrqs}). For the
intermediate temperatures the BS correlation is a little less than the
QS correlation due to the contributions from lighter strange mesons in
the latter. In the HRG model, these two correlations again keep on
increasing indefinitely as higher and higher hadronic states are getting
excited. The behavior in the lattice QCD data as well as the PNJL model
is as it should be in a theory that have partons in the high temperature
phase. This can be understood by noting that at low temperatures the
correlators are small due to the large hadronic masses. They will
increase with temperature, and in the partonic phase the correlation
saturates as there is only one species of quarks containing both
strangeness and baryon number or strangeness and electric charge.  

\par
As in the case of $c_2^S$, the BS and QS correlators in the PNJL-HRG
model underestimate the lattice QCD data in the intermediate range of
temperatures possibly due to the slowly decreasing strange quark mass in
the PNJL model. Therefore it seems that this deviation would pervade all
other observables related to strangeness. A reparametrization of the NJL
part of the model may be able to address this issue and will be
discussed elsewhere.

\section{Conclusion}
\label{sc.conclusion}

\par
In this study we discussed a scheme to address the inadequacy of the
PNJL model in describing the hadronic state of matter as pointed out in
our earlier work in~\cite{repara}. A straightforward approach would be
to add the hadronic contribution from the HRG model.  However a simple
addition of HRG model to PNJL model would have led to overcounting the
degrees of freedom. An interpolating function was therefore used in line
with some earlier studies~\cite{Albright:2014gva, Albright15,
Albright16, iq-hrg1, iq-hrg2}, to smoothly switch between the hadronic
and partonic matter. In general this function would be dependent on
temperature and chemical potentials.  Since we have discussed various
thermodynamic observables vis-a-vis lattice QCD data at zero chemical
potentials, we have considered only a temperature dependent switching
function.

\par
In the earlier studies with the switching function, its temperature and
chemical potential derivatives were considered for computing observables
from thermodynamic relations. However we considered the switching
function to be independent of the observable chosen. Otherwise the
various derivatives of the switching function may be obtained by fitting
with as many number of observables from lattice QCD see
e.g.~\cite{iq-hrg2}.  This would increase the number of free parameters
in the model, thereby reducing its predictive power. Instead we
emphasized on the better agreement of the PNJL model and lattice QCD
data above the crossover region. Therefore a considerable part of the
equation of state in continuum lattice QCD data, both below and above
the crossover temperature, was well described by HRG and PNJL models
respectively.  Hence we only needed to interpolate in the crossover
region using the switching function. Thereafter all other observables
were obtained from thermodynamic relations acting on the individual
pressure of the two models weighed with the switching function.  For
completeness however we have made a comparative study of observables
related to the equation of state by both including and excluding these
derivatives.  All the thermodynamic quantities computed in the PNJL-HRG
model reproduced the lattice QCD data quite well.

\par
Once the switching function was fixed, we computed the predicted
behavior of various fluctuations and correlations of conserved charges.
We found that the fluctuations of baryon number and electric charge
computed in the PNJL-HRG model is in good quantitative agreement with
the lattice QCD data. However the strangeness fluctuations in the PNJL
model was somewhat different from the lattice QCD data around the
crossover region. As a result the PNJL-HRG model did not reproduce
lattice QCD data satisfactorily. We emphasize that this disagreement is
not due to the switching function but rather the inadequacy in the PNJL
model. We argued that since in the PNJL model the constituent mass of
the strange quarks does not decrease fast enough with the rising
temperature, there is a departure of fluctuations of strangeness from
the lattice QCD data.  It naturally followed that none of the leading
order correlators in the PNJL-HRG model could reproduce lattice QCD data
as well. 

\par
Finally we would like to conclude that the scheme of introducing the HRG
model to the PNJL model using the switching function seems to have
worked well. Discrepancies if any are only due to difference between
PNJL model and lattice QCD data in the relevant temperature ranges.
Further for treading into the non-zero chemical potential regions we may
have to consider a chemical potential dependence of the switching
function. We hope to address these issues elsewhere.

\section*{Acknowledgement}
The authors would like to thank Council for Scientific and Industrial
Research (CSIR), Department of Science and Technology (DST), Department
of Atomic Energy (DAE) and Board of Research for Nuclear Sciences (BRNS)
for financial support.



\begin{thebibliography}{99}
\bibitem{Karsch1}
H.-T. Ding, F. Karsch and S. Mukherjee,
in {\it Quark Gluon Plasma 5}, edited by 
X.N. Wang, (World Scientific Publishing, 2016)~
[arXiv: 1504.05274 [hep-lat] (2015)].

\bibitem{Boyd}
G. Boyd, J. Engels, F. Karsch, E. Laermann, C. legeland, M. Lugermeier
and B. Peterson, 
Nucl. Phys. B {\bf 469}, 419 (1996).

\bibitem{Engels}
J. Engels, O. Kaczmarek, F. Karsch and E. Laermann,
Nucl. Phys. B {\bf 558}, 307 (1999).

\bibitem{Katz}
Z. Fodor and S. D. Katz,
Phys. Lett. B {\bf 534}, 87 (2002).

\bibitem{Hands}
C. R. Allton, S. Ejiri, S. J. Hands, O. Kaczmarek, F.  Karsch,
E. Laermann Ch. Schmidt and L. Scorzato,
Phys. Rev. D {\bf 66}, 074507 (2002).

\bibitem{Szabo}
Z. Fodor, S. D. Katz and K. K. Szabo,
Phys. Lett. B {\bf 568}, 73 (2003).

\bibitem{Laermann}
C. R. Allton, S. Ejiri, S. J. Hands, O. Kaczmarek, F.  Karsch,
E. Laermann and Ch. Schmidt, 
Phys. Rev. D {\bf 68}, 014507 (2003).

\bibitem{Philipsen1}
P. de Forcrand and O. Philipsen,
Nucl. Phys. B {\bf 673}, 170 (2003).

\bibitem{Aoki}
Y. Aoki, Z. Fodor, S. D. Katz and K. K. Szabo, 
Phys. Lett. B {\bf 643}, 46 (2006).

\bibitem{Aoki1}
Y. Aoki, G. Endrodi, Z. Fodor, S. D. Katz and K. K. Szabo,
Nature {\bf 443}, 675 (2006).

\bibitem{yaoki}
Y. Aoki, S. Borsanyi, S. Durr, Z. Fodor, S. D. Katz, S. Krieg and
K. K. Szabo,
J. High Energy Phys. {\bf 06}, 088 (2009).

\bibitem{Katz1}
S. Borsanyi, Z. Fodor, S. D. Katz, S. Krieg, C. Ratti and K. K. Szabo, 
Phys. Rev. Lett. {\bf 111}, 062005 (2013).

\bibitem{Bazavov12}
A. Bazavov $et~al.$ (HotQCD collaboration),
Phys. Rev. D {\bf 85}, 054503 (2012).

\bibitem{Bazavov14}
A. Bazavov $et~al.$ (HotQCD collaboration),
Phys. Rev. D {\bf 90}, 094503 (2014).

\bibitem{Borsanyi14}
S. Borsanyi, Z. Fodor, C. Hoelbling, S. D. Katz, S. Krieg and
K. K. Szabo,
Phys. Lett. B {\bf 730}, 99 (2014).

\bibitem{Hatta}
Y. Hatta and M. A. Stephanov,
Phys. Rev. Lett. {\bf 91}, 102003 (2003). 
$Erratum$ {\bf 91}, 129901 (2003).

\bibitem{Ejiri}
S. Ejiri, F. Karsch and K. Redlich,
Phys. Lett. B {\bf 633}, 275 (2006).

\bibitem{Stephanov}
M. A. Stephanov,
Phys. Rev. Lett. {\bf 102}, 032301 (2009).

\bibitem{Asakawa:1989bq}
M. Asakawa and K. Yazaki,
Nucl. Phys. A {\bf 504}, 668 (1989).

\bibitem{Ejiri:2008xt}
S. Ejiri,
Phys. Rev. D {\bf 78}, 074507 (2008).

\bibitem{Bowman:2008kc}
E. S. Bowman and J. I. Kapusta,
Phys. Rev. C {\bf 79}, 015202 (2009).

\bibitem{Halasz:1998qr}
A. M. Halasz, A. D. Jackson, R. E. Shrock, M. A. Stephanov
and J. J. M. Verbaarschot,
Phys. Rev. D {\bf 58}, 096007 (1998).

\bibitem{Fodor:2004nz}
Z. Fodor and S. D. Katz,
J. High Energy Phys. {\bf 04}, 050 (2004).

\bibitem{Gavai:2004sd}
R. V. Gavai and S. Gupta,
Phys. Rev. D {\bf 71}, 114014 (2005).

\bibitem{Stephanov:2004wx}
M. A. Stephanov,
Int. J. Mod. Phys. A {\bf 20}, 4387 (2005).

\bibitem{Finitemu2}
S. Gottlieb, W. Liu, D. Toussaint, R. L. Renken, and R. L. Sugar,
Phys. Rev. Lett. {\bf 59} 2247 (1987);
S. Choe $et~al.$ (QCD-TARO Collaboration),
Phys. Rev. D {\bf 65} 054501 (2002).

\bibitem{Gavai:2001ie} 
R.~V.~Gavai, S.~Gupta and P.~Majumdar,
Phys. Rev. D {\bf 65}, 054506 (2002).

\bibitem{Finitemu1}
M.P. Lombardo and M. d'Elia,
Phys. Rev. D {\bf 67} 014505 (2003).

\bibitem{Gavai1}
R. V. Gavai and S. Gupta,
Phys. Rev. D {\bf 68}, 034506 (2003).

\bibitem{Finitemu3}
S. Gupta and R. Ray,
Phys. Rev. D {\bf 70} 114015 (2004);~
R. Gavai, S. Gupta and R. Ray,
Prog. Theor. Phys. Suppl. {\bf 153} 270 (2004);~
R. V. Gavai and S. Gupta,
Phys. Rev. D {\bf 72} 054006 (2005).

\bibitem{Allton1}
C. Allton, M. Doring, S. Ejiri, S. Hands, O. Kaczmarek, F. Karsch, E.
Laermann, and K. Redlich,
Phys. Rev. D {\bf 71}, 054508 (2005).

\bibitem{Bernard:2004je} 
C.~Bernard {\it et al.} (MILC Collaboration),
Phys. Rev. D {\bf 71}, 034504 (2005).
  
\bibitem{Bernard:2007nm} 
C.~Bernard, C.~E.~DeTar, L.~Levkova, S.~Gottlieb, U.~M.~Heller,
J.~E.~Hetrick, R.~Sugar and D.~Toussaint,
Phys. Rev. D {\bf 77}, 014503 (2008).

\bibitem{Cheng08}
M. Cheng $et. al.$,
Phys. Rev. D {\bf 77}, 014511 (2008).

\bibitem{Kaczmarek}
O. Kaczmarek, F. Karsch, E. Laermann, C. Miao, S. Mukherjee,
P. Petreczky, C. Schmidt, W. Soeldner, and W. Unger,
Phys. Rev. D {\bf 83}, 014504 (2011).

\bibitem{Endrodi1}
G. Endrodi, Z. Fodor, S. D. Katz, K. K. Szabo,
J. High Energy Phys. {\bf 04} 001 (2011).

\bibitem{Endrodi2}
S. Borsanyi, G. Endrodi, Z. Fodor, S. D. Katz, S. Krieg, C. Ratti, and
K. K. Szabo, 
J. High Energy Phys. {\bf 08}, 053 (2012).

\bibitem{YNambu}
Y. Nambu, and G. Jona-Lasinio,
Phys. Rev. {\bf 122}, 345 (1961);
{\bf 124}, 246 (1961).

\bibitem{Hatsuda2}
T. Kunihiro and T. Hatsuda,
Phys. Lett. B {\bf 206}, 385 (1988).

\bibitem{Vogl}
U. Vogl and W. Weise,
Prog. Part. Nucl. Phys. {\bf 27}, 195 (1991). 

\bibitem{Klevansky}
S. P. Klevansky,
Rev. Mod. Phys. {\bf 64}, 649 (1992).

\bibitem{Hatsuda1}
T. Hatsuda and T. Kunihiro,
Phys. Rep. {\bf 247}, 221 (1994).

\bibitem{Buballa1}
M. Buballa,
Phys. Rep. {\bf 407}, 205 (2005).

\bibitem{Barducci}
A. Barducci, R. Casalbuoni, G. Pettini, and L. Ravagli, 
Phys. Rev. D {\bf 72}, 056002 (2005).

\bibitem{Ogilvie}
P. N. Meisinger and M. C. Ogilvie,
Phys. Lett. {\bf B379}, 163 (1996);
Nucl. Phys. B, Proc. Suppl. {\bf 47}, 519 (1996).

\bibitem{Fukushima}
K. Fukushima,
Phys. Lett. B {\bf 591}, 277 (2004).

\bibitem{Ratti}
C. Ratti, M. A. Thaler, and W. Weise,
Phys. Rev. D {\bf 73}, 014019 (2006).

\bibitem{Ray}
S. K. Ghosh, T. K. Mukherjee, M. G. Mustafa, and R. Ray, 
Phys. Rev. D {\bf 73}, 114007 (2006).

\bibitem{Mukherjee}
S. Mukherjee, M. G. Mustafa, and R. Ray,
Phys. Rev. D {\bf 75}, 094015 (2007).

\bibitem{Robner}
C. Ratti, S. Robner, and W. Weise,
Phys. Lett. B {\bf 649}, 57 (2007).

\bibitem{Ghosh}
S. K. Ghosh, T. K. Mukherjee, M. G. Mustafa, and R. Ray, 
Phys. Rev. D {\bf 77}, 094024 (2008).

\bibitem{Datta}
P. Deb, A. Bhattacharyya, S. Datta, and S. K. Ghosh
Phys. Rev. C {\bf 79}, 055208 (2009).

\bibitem{Fu}
W.-j. Fu, Y.-x. Liu, and Y.-L. Wu,
Phys. Rev. D {\bf 81}, 014028 (2010).

\bibitem{Wu}
W.-j Fu and Y.-l. Wu,
Phys. Rev. D {\bf 82}, 074013 (2010).

\bibitem{Bhattacharyya}
A. Bhattacharyya, P. Deb, S. K. Ghosh, and R. Ray,
Phys. Rev. D {\bf 82}, 014021 (2010).

\bibitem{Deb}
A. Bhattacharyya, P. Deb, A. Lahiri, and R. Ray,
Phys. Rev. D {\bf 82}, 114028 (2010).

\bibitem{Lahiri}
A. Bhattacharyya, P. Deb, A. Lahiri, and R. Ray,
Phys. Rev. D {\bf 83}, 014011 (2011).

\bibitem{Osipov}
A. A. Osipov, B. Hiller, and J. da Provid$\hat{e}$ncia, 
Phys. Lett. B {\bf 634}, 48 (2006).

\bibitem{Hiller}
A. A. Osipov, B. Hiller, A. H. Blin, and J. da Provid$\hat{e}$ncia, 
Ann. Phys. (N.Y.) 322, 2021 (2007).

\bibitem{Moreira}
A. A. Osipov, B. Hiller, J. Moreira, A. H. Blin, and
J. da Provid$\hat{e}$ncia,
Phys. Lett. B {\bf 646}, 91 (2007).

\bibitem{Blin}
B. Hiller, J. Moreira, A. A. Osipov, and A. H. Blin,
Phys. Rev. D {\bf 81}, 116005 (2010).

\bibitem{Majumder}
A. Bhattacharyya, S. K. Ghosh, S. Majumder, and R. Ray,
Phys. Rev. D {\bf 86}, 096006 (2012).

\bibitem{Mustafa}
C. A. Islam, R. Abir, M. G. Mustafa, R. Ray, and S. K. Ghosh,
J. Phys. G {\bf 41}, 025001 (2014).

\bibitem{Raha}
A. Bhattacharyya, S. K. Ghosh, A. Lahiri, S. Majumder, S. Raha, and
R. Ray, 
Phys. Rev. C {\bf 89}, 064905 (2014).

\bibitem{Sur}
A. Bhattacharyya, P. Deb, S. K. Ghosh, R. Ray, and S. Sur,  
Phys. Rev. D {\bf 87}, 054009 (2013).

\bibitem{Bhatta}
A. Bhattacharyya, R. Ray, and S. Sur,
Phys. Rev. D {\bf 91(R)}, 051501 (2015).

\bibitem{Kinkar}
A. Bhattacharyya, S. Das, S. K. Ghosh, S. Raha, R. Ray, K. Saha,
and S. Upadhaya, 
arXiv:1212.6010 [hep-ph] (2012).

\bibitem{Anirban}
S. K. Ghosh, A. Lahiri, S. Majumder, M. G. Mustafa, S. Raha and R. Ray, 
Phys. Rev. D {\bf 90}, 054030 (2014).

\bibitem{Redlich}
C. Sasaki, and K. Redlich,
Nucl. Phys. {\bf A832} (2010) 62-75.

\bibitem{Sabyasachi}
S. Ghosh, A. Lahiri, S. Majumder, R. Ray, and S. K. Ghosh, 
Phys. Rev. C {\bf 88}, 068201 (2013).

\bibitem{Marty}
R. Marty, E. Bratkovskaya, W. Cassing, J. Aichelin, and H. Berrehrah, 
Phys. Rev. C {\bf 88}, 045204 (2013).

\bibitem{Weise}
R. Lang, and W. Weise, Eur. Phys. J. A {\bf 50} (2014) 63.

\bibitem{Shi-Song}
X. Shi-Song, G. Pan-Pan, Z. Le, and H. De-Fu, 
Chinese Phys. C {\bf 38}, 054101 (2014).

\bibitem{Sudipa}
S. K. Ghosh, S. Raha, R. Ray, K. Saha, and S. Upadhaya, 
Phys. Rev. D {\bf 91}, 054005 (2015).

\bibitem{Saha}
K. Saha, S. Upadhaya and S. Ghosh,
Mod. Phys. Lett. A {\bf 32}, 1750018 (2017).

\bibitem{Kaiser}
R. Lang, N. Kaiser, and W. Weise,
Eur. Phys. J. A {\bf 51} (2015) 127.

\bibitem{Das}
S. Ghosh, S. K. Das, V. Greco, S. Sarkar, and J. Alam, 
Phys. Rev. D {\bf 90}, 054018 (2014).

\bibitem{Krein}
S. Ghosh, G. Krein, and S. Sarkar,
Phys. Rev. C {\bf 89}, 045201 (2014).

\bibitem{Kadam}
G. P. Kadam and H. Mishra,
Phys. Rev. C {\bf 92}, 035203 (2015).

\bibitem{Mohanty}
S. Ghosh, S. Chatterjee, and B. Mohanty,
Phys. Rev. C {\bf 94}, 045208 (2016).

\bibitem{Claudia}
S. Robner, C. Ratti, and W. Weise,
Phys. Rev. D {\bf 75}, 034007 (2007).

\bibitem{Friman}
C. Sasaki, B. Friman, and K. Redlich,
Phys. Rev. D {\bf 75}, 074013 (2007).

\bibitem{Fuku}
K. Fukushima,
Phys. Rev. D {\bf 77}, 114028 (2008);
{\bf 78}, 039902(E) (2008).

\bibitem{Kahara}
T. Kahara and K. Tuominen,
Phys. Rev. D {\bf 78}, 034015 (2008).

\bibitem{Zhang}
W. Fu, Z. Zhang, and Y. Liu,
Phys. Rev. D {\bf 77}, 014006 (2008).

\bibitem{Ruivo}
P. Costa, M. C. Ruivo, and C. A. de Sousa,
Phys. Rev. D {\bf 77}, 096001 (2008).

\bibitem{Kashiwa}
K. Kashiwa, H. Kouno, M. Matsuzaki, and M. Yahiro,
Phys. Lett. B {\bf 662}, 26 (2008).

\bibitem{Buballa}
M. Buballa, A. G. Grunfeld, A. E. Radzhabov, and D. Scheffler, 
Prog. Part. Nucl. Phys. {\bf62}, 365 (2009).

\bibitem{Hansen}
P. Costa, H. Hansen, M. C. Ruivo, and C. A. de Sousa, 
Phys. Rev. D {\bf 81}, 016007 (2010).

\bibitem{Lourenco}
O. Lourenco, M. Dutra, A. Delfino, and M. Malheiro, 
Phys. Rev. D {\bf 84}, 125034 (2011).

\bibitem{Inagaki}
T. Inagaki, D. Kimura, H. Kohyama, and A. Kvinikhidze,
Phys. Rev. D {\bf 86}, 116013 (2012).

\bibitem{Friesen}
A. V. Friesen, Y. U. L. Kalinovsky, and V. D. Toneev,
Int. J. Mod. Phys. A {\bf 27}, 1250013 (2012).

\bibitem{Sakai}
Y. Sakai, K. Kashiwa, H. Kouno, and M. Yahiro,
Phys. Rev. D {\bf 77}, 051901(R) (2008);
78, 036001(E) (2008).

\bibitem{Yahiro}
Y. Sakai, H. Kouno, and M. Yahiro,
J. Phys. G {\bf 37}, 105007 (2010).

\bibitem{Morito}
K. Morita, V. Skokov, B. Friman, and K. Redlich,
Phys. Rev. D {\bf 84}, 076009 (2011).

\bibitem{Salcedo}
E. Megias, E. R. Arriola, and L. L. Salcedo,
J. High Energy Phys. {\bf 01} (2006) 73.

\bibitem{Sal}
E. Megias, E. R. Arriola, and L. L. Salcedo,
Phys. Rev. Lett. {\bf 109}, 151601 (2012).

\bibitem{Salc}
E. Megias, E. R. Arriola, and L. L. Salcedo,
Phys. Rev. D {\bf 89}, 076006 (2014).

\bibitem{Islam}
C. A. Islam, R. Abir, M. G. Mustafa, R. Ray, and S. K. Ghosh,
J. Phys. G {\bf 41}, 025001 (2014).

\bibitem{Meg}
E. Megias, E. R. Arriola, and L. L. Salcedo,
Phys. Rev. D {\bf 74}, 065005 (2006).

\bibitem{Megi}
E. Megias, E. R. Arriola, and L. L. Salcedo,
Phys. Rev. D {\bf 74}, 114014 (2006).

\bibitem{Tsai}
H.-M. Tsai and B. Muller,
J. Phys. G {\bf 36}, 075101 (2009).

\bibitem{braun}
J. Braun, L. M. Haas, F. Marhauser and J. M. Pawlowski,
Phys. Rev. Lett. {\bf 106}, 022002 (2011);
L. M. Haas, R. Stiele, J. Braun, J. M. Pawlowski and
J. Schaffner-Bielich,
Phys. Rev. D {\bf 87}, 076004 (2013).

\bibitem{rincon}
J. M. Torres-Rincon and J. Aichelin,
arXiv:1601.01706 [nucl-th] (2016);
arXiv:1704.07858 [nucl-th] (2017).

\bibitem{Bazavov:2012jq} 
A.~Bazavov {\it et al.} (HotQCD Collaboration),
Phys. Rev. D {\bf 86}, 034509 (2012)

\bibitem{Borsanyi:2011sw} 
S.~Borsanyi, Z.~Fodor, S.~D.~Katz, S.~Krieg, C.~Ratti and K.~Szabo,
J. High Energy Phys. {\bf 1201}, 138 (2012).
  
\bibitem{Bazavov1213}
A. Bazavov {\it et al.},
Phys. Rev. Lett. {\bf 109}, 192302 (2012);
C.  Schmidt (for the BNL-Bielefeld Collaboration),
Proc. Sci., ConfinementX2012 (2012) 187;
A. Bazavov {\it et al.},
Phys. Rev. Lett. {\bf 111}, 082301 (2013).

\bibitem{Bazavov14a}
A. Bazavov {\it et al.},
Phys. Rev. Lett. {\bf 113}, 072001 (2014).

\bibitem{repara}
A. Bhattacharyya, S. K. Ghosh, S. Maity, S. Raha, R. Ray, K. Saha
and Sudipa Upadhaya,
Phys. Rev. D {\bf 95}, 054005 (2017).

\bibitem{HRG_Braun-Munzinger}
P. Braun-Munzinger, K. Redlich and J. Stachel,
in {\it Quark Gluon Plasma 3}, edited by 
R.C. Hwa and X.N. Wang, (World Scientific Publishing, 2004).


\bibitem{BraunMunzinger:1994xr} 
P.~Braun-Munzinger, J.~Stachel, J.~P.~Wessels and N.~Xu,
Phys. Lett. B {\bf 344}, 43 (1995).
  
\bibitem{9603004_Cleymans}
J. Cleymans, D. Elliott, H. Satz and R. L. Thews,
Z. Phys. C {\bf 74}, 319 (1997).

\bibitem{PLB465_Braun-Munzinge}
P. Braun-Munzinger, I. Heppe and J. Stachel,
Phys.  Lett. B {\bf 465}, 15 (1999).

\bibitem{PRC60_054908_Cleymans}
J. Cleymans and K. Redlich, 
Phys. Rev. C {\bf 60}, 054908 (1999).

\bibitem{PLB518_Braun-Munzinger}
P. Braun-Munzinger, D. Magestro, K. Redlich and J. Stachel, 
Phys. Lett. B {\bf 518}, 41 (2001).

\bibitem{Xu:2001zj}
N. Xu and M. Kaneta,
Nucl. Phys. A {\bf 698}, 306 (2002).

\bibitem{PRC73_Becattini}
F. Becattini, J. Manninen and M. Gazdzicki,
Phys. Rev. C {\bf 73}, 044905 (2006).

\bibitem{NPA772_Andronic}
A. Andronic, P. Braun-Munzinger and J. Stachel, 
Nucl. Phys. A {\bf 772}, 167 (2006).

\bibitem{Cleymans:2005xv}
J. Cleymans, H. Oeschler, K. Redlich and S. Wheaton,
Phys. Rev. C {\bf 73}, 034905 (2006).

\bibitem{Andronic:2008gu} 
A.~Andronic, P.~Braun-Munzinger and J.~Stachel,
Phys.\ Lett.\ B {\bf 673}, 142 (2009).
  
\bibitem{Andronic:2009jd}
A. Andronic, P. Braun-Munzinger and J. Stachel,
Nucl. Phys. A {\bf 834}, 237C (2010).

\bibitem{Karsch:2010ck} 
F.~Karsch and K.~Redlich,
Phys. Lett. B {\bf 695}, 136 (2011).

\bibitem{Chatterjee:2015fua}
S. Chatterjee, S. Das, L. Kumar, D. Mishra, B. Mohanty, R. Sahoo and
N. Sharma,
Adv. High Energy Phys. {\bf 2015}, 349013 (2015).

\bibitem{Karsch:2003zq}
F. Karsch, K. Redlich and A. Tawfik
Phys. Lett. B {\bf 571}, 67 (2003).

\bibitem{Tawfik:2004sw}
A. Tawfik,
Phys. Rev. D {\bf 71}, 054502 (2005).

\bibitem{Andronic:2012ut} 
A.~Andronic, P.~Braun-Munzinger, J.~Stachel and M.~Winn,
Phys. Lett. B {\bf 718}, 80 (2012).
  
\bibitem{Bhattacharyya:2013oya} 
A.~Bhattacharyya, S.~Das, S.~K.~Ghosh, R.~Ray and S.~Samanta,
Phys. Rev. C {\bf 90}, 034909 (2014).

\bibitem{dashen}
R. Dashen, S.-K. Ma, and H. J. Bernstein, Phys. Rev. 187, 345
(1969); R. Dashen and S. K. Ma, Phys. Rev. A 4, 700 (1971).

\bibitem{hagedron:1980}
R.~ Hagedorn, J.~ Rafelski.,
Phys. Lett. B {\bf 97}, 136 (1980).
  
\bibitem{ZPC51_Rischke}
D.H. Rischke, M. I. Gorenstein, H. St$\ddot{o}$cker and W. Greiner,
Z. Phys. C {\bf 51}, 485 (1991).

\bibitem{Cleymans:1992jz} 
J.~Cleymans, M.~I.~Gorenstein, J.~Stalnacke and E.~Suhonen,
Phys. Scripta {\bf 48}, 277 (1993).
  
\bibitem{Singh:1991np} 
C.~P.~Singh, B.~K.~Patra and K.~K.~Singh,
Phys. Lett. B {\bf 387}, 680 (1996).
  
\bibitem{Yen:1997rv} 
G.~D.~Yen, M.~I.~Gorenstein, W.~Greiner and S.~N.~Yang,
Phys. Rev. C {\bf 56}, 2210 (1997).
  
\bibitem{PRC77_Gorenstein}
M. I. Gorenstein, M. Hauer, and O. N. Moroz, 
Phys. Rev. C {\bf 77}, 024911 (2008).

\bibitem{PRC85_Fu}
J. Fu, 
Phys. Rev. C {\bf 85}, 064905 (2012).

\bibitem{begunprc88}
V. V. Begun, M. Ga\'{z}dzicki, and M. I. Gorenstein, 
Phys. Rev. C {\bf 88}, 024902 (2013).

\bibitem{PLB722_Fu}
J. Fu,
Phys. Lett. B {\bf 722}, 144 (2013).

\bibitem{PRC88_Tawfik}
A. Tawfik, 
Phys. Rev. C {\bf 88}, 035203 (2013).

\bibitem{hama}
Y. Hama, T. Kodama, and O. Socolowski,
Braz. J. Phys. {\bf 35}, 24 (2005).  

\bibitem{werner}
K. Werner, Iu. Karpenko, T. Pierog, M. Bleicher and K.  Mikhailov, 
Phys. Rev. C {\bf 82} , 044904 (2010).

\bibitem{satarov}
A. V. Merdeev, L. M. Satarov, and I. N. Mishustin,
Phys. Rev. C {\bf 84}, 014907 (2011).

\bibitem{Garg:2013ata} 
P.~Garg, D.~K.~Mishra, P.~K.~Netrakanti, B.~Mohanty, A.~K.~Mohanty,
B.~K.~Singh and N.~Xu,
Phys. Lett. B {\bf 726}, 691 (2013).
   
\bibitem{Bhattacharyya:2015zka} 
A.~Bhattacharyya, R.~Ray, S.~Samanta and S.~Sur,
Phys. Rev. C {\bf 91}, 041901 (2015).
  
\bibitem{hrgmag}
A. Bhattacharyya, S. K. Ghosh, R. Ray and S. Samanta,
Europhys. Lett. {\bf 115}, 62003 (2016).

\bibitem{hrgcentral}
R. P. Adak, S. Das, S. K. Ghosh, R. Ray and S. Samanta,
Phys. Rev. C {\bf 96}, 014902 (2017).

\bibitem{Albright:2014gva} 
M.~Albright, J.~Kapusta and C.~Young,
Phys. Rev. C {\bf 90}, 024915 (2014).

\bibitem{Albright15}
M. Albright, J. Kapusta and C. Young, 
Phys. Rev. C {\bf 92}, 044904 (2015).

\bibitem{Albright16}
J. Kapusta, M. Albright and C. Young, 
Eur. Phys. J. A {\bf 52}, 250 (2016).

\bibitem{iq-hrg1}
A. Miyahara, Y. Torigoe, H. Kouno and M. Yahiro,
Phys. Rev. D {\bf 94}, 016003 (2016).

\bibitem{iq-hrg2}
A. Miyahara, M. Ishii, H. Kouno and M. Yahiro,
arXiv:1704.06432 [hep-ph] (2017).

\bibitem{ciminale}
M. Ciminale, R. Gatto, N. D. Ippolito, G. Nardulli, and M. Ruggieri,
Phys. Rev. D {\bf 77}, 054023 (2008).

\bibitem{Shao}
G. Y. Shao, Z. D. Tang, M. Di Toro, M. Colonna, X. Y. Gao, and N. Gao,
Phys. Rev. D {\bf 94}, 014008 (2008).

\bibitem{Tang}
G. Y. Shao, Z. D. Tang, M. Di Toro, M. Colonna, X. Y. Gao, N. Gao, and
Y. L. Zhao,
Phys. Rev. D {\bf 92}, 114027 (2015).

\bibitem{Haque}
C. A. Islam, S. Majumder, N. Haque, and M. G. Mustafa,
J. High Energy Phys. {\bf 1502} (2015) 011.

\bibitem{Peixoto}
S. Ghosh, T. C. Peixoto, V. Roy, F. E. Serna, and G. Krein,
Phys. Rev. D {\bf 93}, 045205 (2016).

\bibitem{Contrera}
G. A. Contrera, A. G. Grunfeld, and D. B. Blaschke,
Phys. of Part. and Nucl. Lett., 2014, Vol 11, No. 4, pp 342-351.

\bibitem{Qin}
X. Xin, S. Qin, and Y. Liu,
Phys. Rev. D {\bf 89}, 094012 (2014).

\bibitem{jweber}
A. Bazavov, N. Brambilla, H. -T. Dong, P. Petreczky, H. -P. Schadler, A.
Vairo, and J. H. Weber,
Phys. Rev. D {\bf 93}, 114502 (2016).

\bibitem{pdg}
K. A. Olive et al. (Particle Data Group),
Chin. Phys. C {\bf 38}, 090001 (2014).
  
\bibitem{Hung95}
C. M. Hung and E.V. Shuryak,
Phys. Rev. Lett. {\bf 75} 4003 (1995).

\bibitem{Koch:2008ia} 
V.~Koch,
Chapter of the book "Relativistic Heavy Ion Physics", R. Stock (Ed.), 
Springer, Heidelberg, 2010, p. 626-652.
(Landolt-Boernstein New Series I, v. 23).
(ISBN: 978-3-642-01538-0, 978-3-642-01539-7 (eBook))
  [arXiv:0810.2520 [nucl-th]]. 

\bibitem{Abelev}
B. Abelev $et~al.$,
Phys. Rev. Lett. {\bf 110}, 152301 (2013).

\bibitem{Shuryak}
M. A. Stephanov, K. Rajagopal and E. V. Shuryak, 
Phys. Rev. Lett. {\bf 81}, 4816 (1998).

\bibitem{Jeon}
S. Jeon and V. Koch,
Phys. Rev. Lett. {\bf 83}, 5435 (1999).

\bibitem{Jeon1}
S. Jeon and V. Koch,
Phys. Rev. Lett. {\bf 85}, 2076 (2000).

\bibitem{Heinz}
M. Asakawa, U. Heinz, and B. Muller, 
Phys. Rev. Lett. {\bf 85}, 2072 (2000).
  
\bibitem{Upadhaya}
A. Bhattacharyya, S. K. Ghosh, R. Ray, K. Saha, and S. Upadhaya,
Europhys. Lett. {\bf 116}, 52001 (2016).

\bibitem{Bellwied1}
R. Bellwied, S. Borsanyi, Z. Fodor, S. D. Katz and C. Ratti,
Phys. Rev. Lett {\bf 111}, 202302 (2013).

\bibitem{Swagato1}
A. Bazavov, H.-T. Ding, P. Hegde, F. Karsch, C. Miao, S.
Mukherjee, P. Petreczky, C. Schmidt, and A. Velytsky, Phys.
Rev. D {\bf 88}, 094021 (2013).

\bibitem{Swagato2}
H.-T. Ding, S. Mukherjee, H. Ohno, P. Petreczky and H.-P. Schadler,
Phys. Rev. D {\bf 92}, 074043 (2015).

\bibitem{Bellwied2}
R. Bellwied, S. Borsanyi, Z. Fodor, S. D. Katz, A. Pasztor,
C. Ratti and K. K. Szabo,
Phys. Rev. D {\bf 92}, 114505 (2015).

\end{thebibliography}
\end{document}